\documentclass[pdflatex,sn-mathphys-num]{sn-jnl}

\usepackage{graphicx}%
\usepackage{multirow}%
\usepackage{amsmath,amssymb,amsfonts}%
\usepackage{amsthm}%
\usepackage{mathrsfs}%
\usepackage[title]{appendix}%
\usepackage{xcolor}%
\usepackage{textcomp}%
\usepackage{manyfoot}%
\usepackage{booktabs}%
\usepackage{algorithm}%
\usepackage{algorithmicx}%
\usepackage{algpseudocode}%
\usepackage{listings}%
\usepackage{cleveref}%
\usepackage{bbm}%
\usepackage{subcaption}
\usepackage{placeins}
\usepackage{footmisc}
\setcitestyle{numbers,sort&compress}

\usepackage{float}

\newfloat{extfigure}{htbp}{loe}
\floatname{extfigure}{Figure}
\newfloat{exttable}{htbp}{loeT}
\floatname{exttable}{Table}
\newfloat{suppfigure}{htbp}{los}
\floatname{suppfigure}{Figure}
\newfloat{supptable}{htbp}{losT}
\floatname{supptable}{Table}
\floatstyle{plain}
\restylefloat{extfigure}
\restylefloat{exttable}
\restylefloat{suppfigure}
\restylefloat{supptable}

\crefname{extfigure}{Extended Data Figure}{Extended Data Figures}%
\Crefname{extfigure}{Extended Data Figure}{Extended Data Figures}%
\crefname{exttable}{Extended Data Table}{Extended Data Tables}%
\Crefname{exttable}{Extended Data Table}{Extended Data Tables}%
\crefname{suppfigure}{Supp. Info. Figure}{Supp. Info. Figures}%
\Crefname{suppfigure}{Supp. Info. Figure}{Supp. Info. Figures}%
\crefname{supptable}{Supp. Info. Table}{Supp. Info. Tables}%
\Crefname{supptable}{Supp. Info. Table}{Supp. Info. Tables}%

\makeatletter
\newcommand{\setextdataanchors}{%
    \renewcommand{\theHsection}{ext.S\arabic{section}}%
}
\makeatother

\newcommand{\beginextdata}{%
    \setcounter{table}{0}%
    \renewcommand{\thetable}{S\arabic{table}}%
    \setcounter{figure}{0}%
    \renewcommand{\thefigure}{S\arabic{figure}}%
    \setcounter{section}{0}%
    \renewcommand{\thesection}{S\arabic{section}}%
        \setextdataanchors%
    \let\origfigure\figure
    \let\endorigfigure\endfigure
    \let\figure\extfigure
    \let\endfigure\endextfigure
    \let\origtable\table
    \let\endorigtable\endtable
    \let\table\exttable
    \let\endtable\endexttable
}

\newcommand{\beginsupplement}{%
    \setcounter{table}{0}%
    \renewcommand{\thetable}{S\arabic{table}}%
    \setcounter{figure}{0}%
    \renewcommand{\thefigure}{S\arabic{figure}}%
    \let\figure\suppfigure
    \let\endfigure\endsuppfigure
    \let\table\supptable
    \let\endtable\endsupptable
}

\newcommand{\tildeXle}[1]{\ensuremath{\tilde{X}^{\leq #1}}}

\newcommand{\E}{\mathbb{E}} % expectation
\newcommand{\N}{\mathbb{N}} 
\newcommand{\R}{\mathbb{R}}

\begin{document}

%%======================================================================%%
%% Title & authors %%
%TC:ignore
\title[Dynamics of antibiotic anomalies in infant microbiome with neural jump ODEs]{Revealing the temporal dynamics of antibiotic anomalies in the infant gut microbiome with neural jump ODEs}

\author[1]{\fnm{Anja} \sur{Adamov}}
\equalcont{* These authors contributed equally to this work.}
\author[2]{\fnm{Markus} \sur{Chardonnet}}
\equalcont{These authors contributed equally to this work.}
\author[2]{\fnm{Florian} \sur{Krach}}
\equalcont{These authors contributed equally to this work.}
\author[2,3]{\fnm{Jakob} \sur{Heiss}}
% \author[4]{\fnm{Alina} \sur{Ofenheimer}}

\author*[2]{\fnm{Josef} \sur{Teichmann}}\email{josef.teichmann@math.ethz.ch}
\author*[1]{\fnm{Nicholas A.} \sur{Bokulich}}\email{nicholas.bokulich@hest.ethz.ch}

\affil[1]{\orgdiv{Department of Health Sciences and Technology}, \orgname{ETH Zurich}, \country{Switzerland}}

\affil[2]{\orgdiv{Department of Mathematics}, \orgname{ETH Zurich}, \country{Switzerland}}

\affil[3]{\orgdiv{Department of Statistics}, \orgname{University of California, Berkeley}, \country{USA}}

%%======================================================================%%
%% Abstract %%
%% limit: up to 150 words, 
%% style: unreferenced
\abstract{
Detecting anomalies in irregularly sampled multi-variate time-series is challenging, especially in data-scarce settings. Here we introduce an anomaly detection framework for irregularly sampled time-series that leverages neural jump ordinary differential equations (NJODEs). The method infers conditional mean and variance trajectories in a fully path dependent way and computes anomaly scores. On synthetic data containing jump, drift, diffusion, and noise anomalies, the framework accurately identifies diverse deviations. Applied to infant gut microbiome trajectories, it delineates the magnitude and persistence of antibiotic-induced disruptions: revealing prolonged anomalies after second antibiotic courses, extended duration treatments, and exposures during the second year of life. We further demonstrate the predictive capabilities of the inferred anomaly scores in accurately predicting antibiotic events and outperforming diversity-based baselines. Our approach accommodates unevenly spaced longitudinal observations, adjusts for static and dynamic covariates, and provides a foundation for inferring microbial anomalies induced by perturbations, offering a translational opportunity to optimize intervention regimens by minimizing microbial disruptions.

% \com{Nature methods "Article" restrictions: main text: \textbf{3'000 words} (max. 5'000 with editorial discretion) excl. Abstract, Methods, references and figure legends; up to \textbf{6 figures and/or tables}.}
% \com{Upload extended data with "link to file" with submission or make repos public}
}

\maketitle

%TC:endignore
%%======================================================================%%
%% Main text %%

%%===============================%%
%% Introduction %%
%% style: referenced text without subheadings
\section{Introduction}\label{intro}

The infant gut microbiota exhibits a gradual succession during the first years of life to reach an adult-like microbial composition \cite{dogra2021, stewart2018}. The trajectory of development has been shown to play a crucial role in the overall health of an infant, with alterations in early gut microbiome development being associated with metabolic \cite{chen2020, blanton2016}, immune \cite{kostic2015, bisgaard2011} and inflammatory disorders \cite{stokholm2018}. Infant gut microbiome maturation is influenced by various exposures, including diet \cite{bokulich2016, stewart2018}, delivery mode \cite{shao2019, bokulich2016, stewart2018} and antibiotics \cite{mcdonnell2021, bokulich2016}. Antibiotic exposures in infants have repeatedly been shown to reduce microbial diversity and richness, and reduce the abundance of specific microbial taxa such as \textit{Bifidobacterium}, \textit{Proteobacteria} and \textit{Lactobacillus} \cite{mcdonnell2021}. Despite these findings, the duration, frequency, and timing effects of antibiotic administration on the infant gut microbiome remain largely unknown, in part due to challenges in modeling longitudinal dynamics of gut microbiota that are characterized by high inter-individual and temporal variability.

Current approaches to study the effects of antibiotics on infant gut microbiome development often involve comparing antibiotic-exposed and unexposed groups by assessing the difference in temporal trajectories of within-sample diversity \cite{yassour2016, bokulich2016} and microbiota-by-age z-scores \cite{subramanian2014, bokulich2016}, or by quantifying the dissimilarities between consecutive samples \cite{yassour2016,vatanen2018}. Association studies have also been employed, correlating occurrence or number of antibiotics exposures prior to sampling with their alpha- \cite{vatanen2018} and beta-diversity values \cite{li2023}, and taxonomic groups \cite{vatanen2016}. The limitations of these approaches include a coarse grouping of samples (e.g., exposed vs. unexposed; before vs. after antibiotics exposures) that potentially obscures signals occurring at particular time points. These approaches do not account for the exact timing of antibiotics administration and fail to quantify the duration of the antibiotics effects on the gut microbiome. Additionally, most studies do not adjust for important covariates, such as delivery mode and dietary changes. Hence, these approaches are limited in their ability to differentiate anomalies, i.e., periods of abnormal development, from regular fluctuations in microbiota composition.

Modeling (ab)normal temporal dynamics in microbiome data, for forecasting-based anomaly detection, is challenging due to irregular temporal sampling schemes and small sample sizes. In machine learning, standard time-series models include recurrent neural networks (RNNs), such as LSTMs \citep{LSTMhochreiter1997long}, that are restricted to modeling regularly observed or imputed data. Transformers \citep{NIPS2017_3f5ee243AttentionIsAllYouNeed} pose a different temporal modeling approach; however, they might not have an appropriate inductive bias, which is particularly important in small data regimes. Alternatively, state space models \citep{SSMgu2021efficiently} require large amounts of training data. By leveraging pre-trained foundational models, TabPFN-TS \citep{hoo2025tabularfoundationmodeltabpfn} or TiRex \citep{auer2025tirexzeroshotforecastinglong} can be employed on new time-series without finetuning. However, they are restricted in requiring a long enough history of a trajectory to make accurate individual predictions.
Methods for anomaly detection include the dissimilarity-based method \citep{OCSVM,DBSCAN,dict_AD,Connectivity_value}, frequency-based methods \citep{frequency_based,periodic_frequency_based}, reconstruction-based methods using RNNs, convolutional neural networks or transformers \citep{RNN_AD_1,CNN_AD_1,Transformer_AD_1}, and graph-based forecasting methods \citep{GTA,GDN,Mutlivariate_TSAD,review_DL_AD}, all of which struggle with limited amounts of data or scarce, irregular observations.

In this study, we aim to address the limitations of existing methods in revealing the temporal dynamics of antibiotic anomalies in the infant microbiome with a novel anomaly framework (\Cref{sec:The anomaly framework design}) that leverages neural jump ordinary differential equations (NJODEs) \cite{herrera2021neural,krach2022optimal,NJODE3,krach2024,heiss2024nonparametricfilteringestimationclassification,ThesisKrach20.500.11850/720717} (see \Cref{appendix:LiteratureAlternativesToNJODE} for a comparison to other methods). 
We first demonstrate the strength of this method to detect many different types of anomalies in synthetic time-series data (\Cref{result:simulate}).
By inferring temporal dynamics of the gut microbiome of healthy, antibiotic-unexposed infants (\Cref{result:application_microbiome}), we manage to accurately describe (\Cref{result:describe1,result:describe2}) and detect (\Cref{result:predict}) antibiotic-induced anomalies in infants that were exposed to antibiotics. Crucially, our framework accommodates irregularly observed data and rigorously adjusts for static and dynamic covariates such as delivery mode and dietary transitions.

%%===============================%%
%% Results %%
%% style: referenced text with subheadings
\section{Results}\label{results}

\subsection{The anomaly detection framework design}\label{sec:The anomaly framework design}
    Our anomaly detection framework consists of a Neural Jump ODE (NJODE) predictive model that estimates the conditional distribution of a target process given past observations and an anomaly detection algorithm that identifies outliers based on the inferred estimator.

\begin{figure}[hb]
    \centering
    \includegraphics[width=\linewidth]{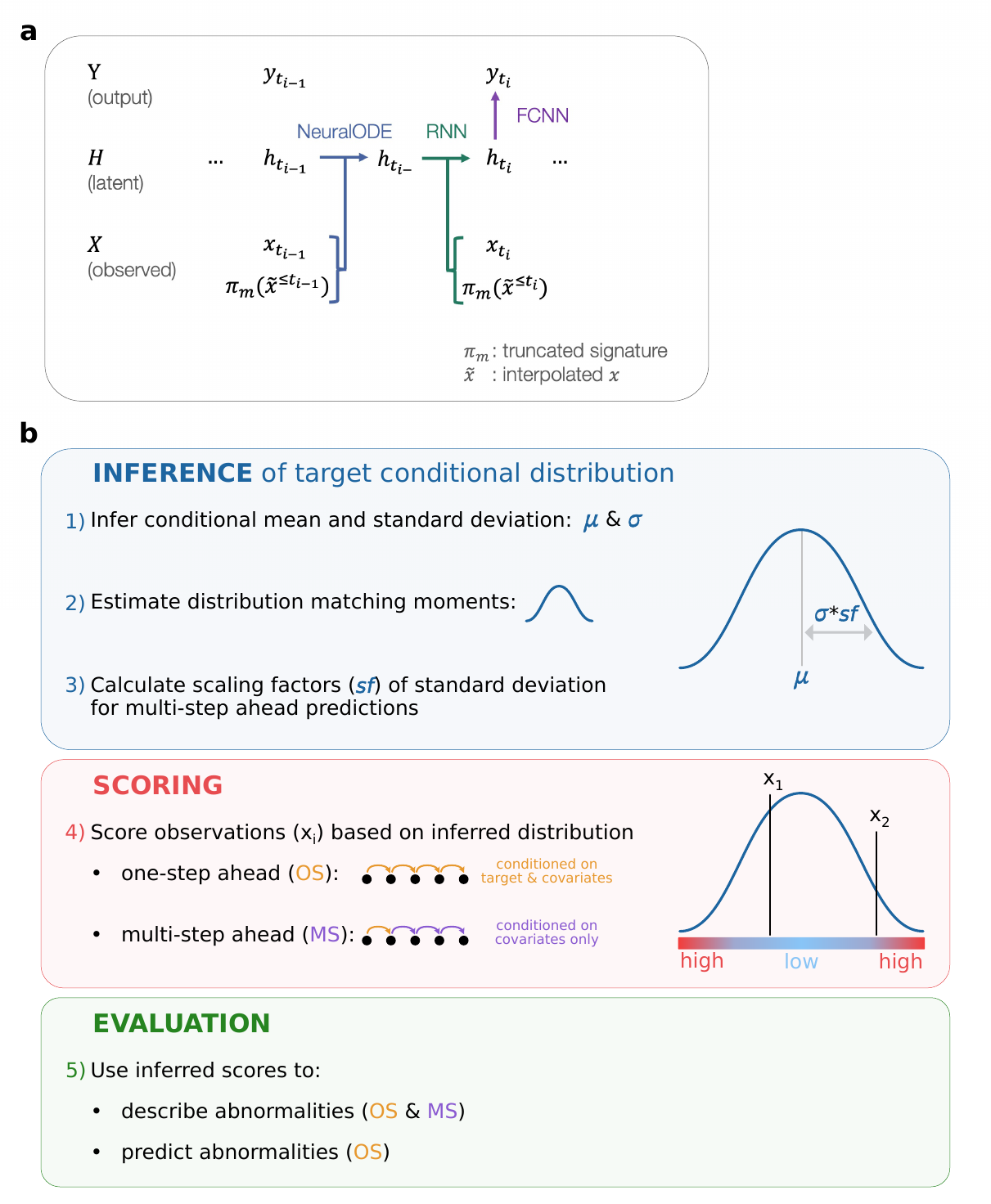}
    \caption{Overview of the anomaly detection framework design.
        \textbf{(a)} Schematic of the three neural network components of the NJODE model for inferring the target $y_{t_i}$, with RNN = recurrent neural network and FCNN = fully connected neural network.
        \textbf{(b)} Description of the three modules of the anomaly detection algorithm.
    }
    \label{fig:schematic}
\end{figure}

    \textbf{NJODE.} The main goal of NJODEs is to learn the dynamics of the optimal prediction of an observed target process $X = (X_t)_{t \in [0,T]}$, which in an $L^2$-sense is given by the conditional expectation. These dynamics can be described by a differential equation ($f_{\theta_1}$ in \eqref{equ:PD-NJ-ODE}), between any two observation times (i.e, whenever the information to condition on is constant). At timestamps where new observations occur, the dynamics jump with $\rho_{\theta_2}$ in \eqref{equ:PD-NJ-ODE}. The complete NJODE model is defined by
    \begin{equation}\label{equ:PD-NJ-ODE}
    \begin{split}
    H_0 &= \rho_{\theta_2}\left(0, 0, \pi_m (0), Z_0 \right), \\
    dH_t &= f_{\theta_1}\left(H_{t-}, t, \tau(t), \pi_m (\tildeXle{\tau(t)} ), Z_{\tau(t)} \right) dt  \\
    & \quad + \left( \rho_{\theta_2}\left( H_{t-}, t, \pi_m (\tildeXle{\tau(t)}), Z_{\tau(t)} \right) - H_{t-} \right) du_t, \\
    Y_t &= g_{\theta_3}(H_t),
    \end{split}
    \end{equation}
    where the model output $Y_t$ corresponds to the conditional expectation of $(X_t, X_t^2)$, which can be transformed to the conditional mean and variance $(\mu_t,\sigma_t^2)$ \citep[see][Sec. 5]{krach2022optimal} and is further used in the next anomaly detection step. $Z_t = (X_t, C_t)$ is a process consisting of the target process $X$ and a (potential) additional covariate process $C$; $H_t$ corresponds to a hidden state, $\pi_m (\tildeXle{\tau(t)} )$ corresponds to a feature transformation of the observed history based on the signature transform (\Cref{sec:signature}), $\tau(t)$ is the last observation time prior to $t$, and $u_t$ counts the number of observations. 
    We represent the unknown functions that appear in the formulation of the ODE~\eqref{equ:PD-NJ-ODE} by neural networks $f_{\theta_1}, \rho_{\theta_2}, g_{\theta_3}$. 
    A schematic overview how the NJODE processes the irregular inputs to generate predictions is given in \Cref{fig:schematic}a.
    For further details on the model see \Cref{sec:Details for the NJODE model}.
    
    \textbf{Anomaly detection.} The anomaly detection algorithm starts with the inferred conditional mean and variance, $\mu_t$ and $\sigma_t^2$, from the NJODE model (step 1 in \Cref{fig:schematic}b). These are then used to match a distribution of a prespecified family by estimating its parameters via the method of moments \citep[see][Sec. 0.2.3]{KrachPhDThesis} (step 2 in \Cref{fig:schematic}b). To account for the increasing epistemic uncertainty with long-term predictions (\Cref{sec:Details for anomaly detection}), we fit scaling factors (SF) s.t.\ the resulting empirical standardized conditional distributions of observations match the theoretical standardized distribution well (step 3 in \Cref{fig:schematic}b, \Cref{sec:Details for the computation of scaling factors}).
    The SFs are used in the computation of the estimated conditional target distribution, which, in turn, is used to calculate p-values of new observations (step 4 in \Cref{fig:schematic}b). A small p-value means that the observed value is unlikely in terms of the estimated conditional distribution, hence potentially anomalous.
    To allow for a better distinction of small p-values, we transform them into anomaly scores with $S = -\ln(p)$, where higher values correspond to larger anomalies.
    %Using the inferred scores, we leverage our framework to describe and detect anomalies after predefined cutoff times, which set the information window for scoring (steps 4 and 5 in \Cref{fig:schematic}b). When the cutoff is set to the most recent observation, we obtain the standard one-step ahead score. To isolate the impact of external events, we place the cutoff at the event time and compute multi-step-ahead scores conditioning only on the covariates for all subsequent observations.
    Using this framework, we can %both compute %one-step-ahead scores (conditioning on all information %$Z^{\leq \tau(t-)}$
    %observed berfore time $t$) or
    compute multi-step-ahead scores, conditioning only on the targets $X$ %$X^{\leq T}$
    observed before a cutoff time $s<t$ and the covariances $C$ %$C^{\leq \tau(t-)}$
    observed before time $t$, (steps 4 and 5 in \Cref{fig:schematic}b). The multi-step-ahead scores estimate the long-term effects of anomalous events since the cutoff time $s$ on $X_t$. %We can place the cutoff time $s$ directly before an antibiotics exposure.

\subsection{Verification of anomaly framework on simulated time-series}\label{result:simulate}
\begin{figure}[hp]
    \centering
    \includegraphics[width=\linewidth]{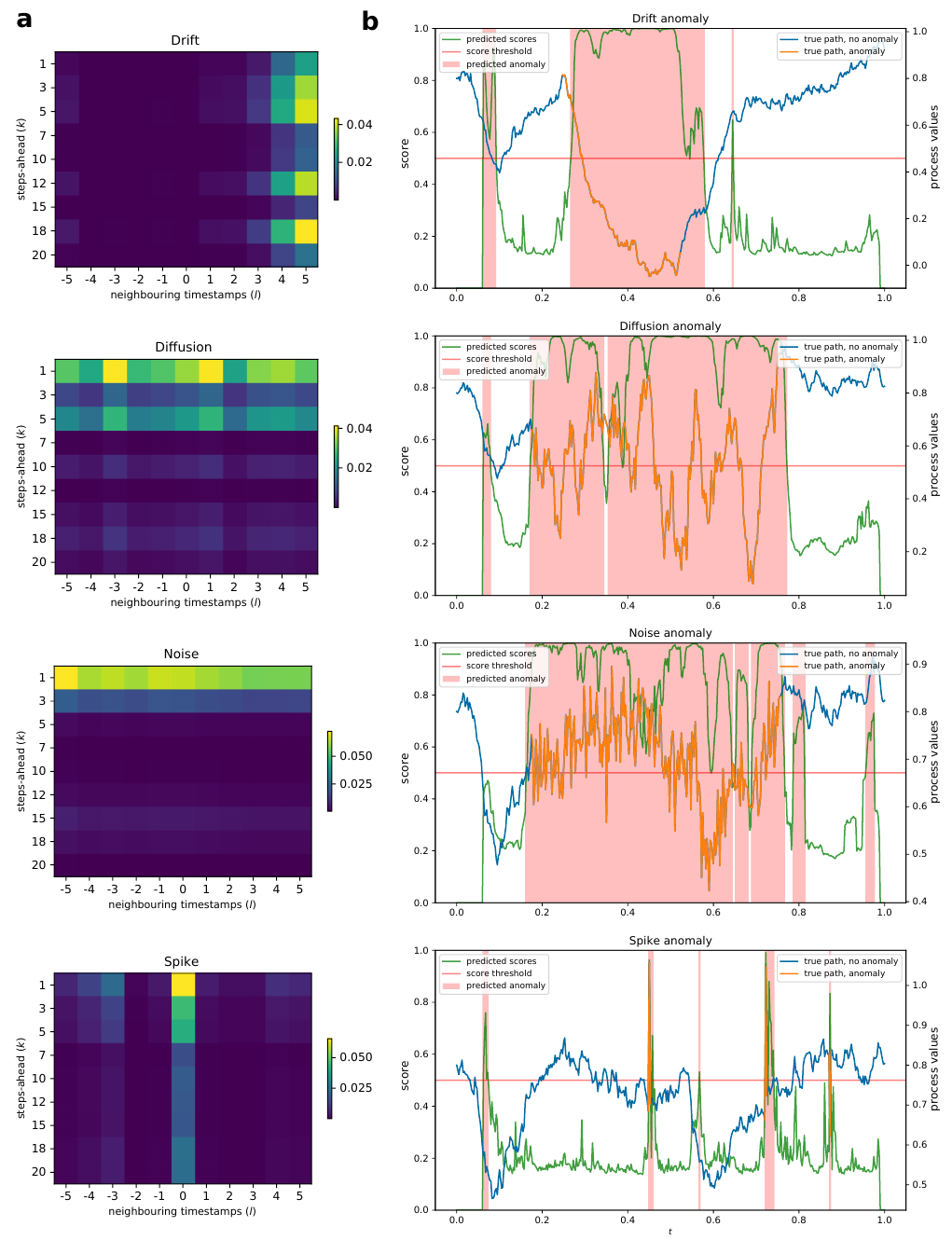}
    \caption{Anomalies detected in simulated time-series.
        \textbf{(a)} Learned score aggregation weights for different anomaly types. The horizontal axis shows the influence of neighboring scores (past observations on the left and future on the right), while the vertical axis shows the influence of different forecasting horizons.
        \textbf{(b)} Example plots of anomaly detection on the different synthetic anomaly type test sets. The ground truth path is colored orange when anomalous and blue otherwise. The predicted aggregated scores are in green, and the red line is the score threshold level of 0.5 to label an observation as anomalous. Predicted anomaly regions are shaded in red.
    }
    \label{fig:simulation_results}
\end{figure}

To verify that our anomaly framework accurately detects anomalies in irregularly sampled time series, we applied it on synthetically generated data.
The NJODE model was trained on a large anomaly-free training set, inferred from a base data model, to learn the baseline dynamics. 
The base model is a diffusion process $X=(X_{t})_{0\leq t\leq T}$, defined as a generalization of an Ornstein-Uhlenbeck process (cf.~\Cref{sec:The synthetic base data model}).
In addition, four \emph{anomalous datasets} were generated, each with a different type of anomaly injected into the base data model. The anomalies include: i) a change of drift, ii) a change of diffusion, iii) added noise, and iv) spikes (\Cref{suppfig:anomaly types} and \Cref{sec:Injected anomalies}).
For the anomalous datasets, we assumed regular dense observations on the equidistant sampling grid with grid size $\delta$, removing the need for scaling factors.
Given the regular dense observations on the anomalous dataset, we computed several scores $S_{t,s}$ at any (grid) time $t$ for multiple cutoff times $s < t$. 
Then we defined the aggregated score at $t$ as the linear combination
\begin{equation*}
    S^{\text{ag}}_t = \sum\limits_{l=-L}^{L}\sum\limits_{k \in \mathcal{K}}w_{lk} S_{ t + \delta l, t + \delta (l-k)}.
\end{equation*}
We learn the aggregation weights $w_{lk}$ in the logistic regression problem to correctly classify the anomaly label $y_t$ at each sampling grid point with stochastic gradient descent on a labeled part of the anomalous datasets (the \emph{aggregation training sets}, cf.~\Cref{sec:Injected anomalies}).
In \Cref{fig:simulation_results}a, we show the learned aggregation weights for the different types of anomalies. 
One key insight is that a spike anomaly at time $t$ can be detected only using scores $S_{t,s}$ at this time $t$ ($l=0$), i.e., the neighboring scores do not contribute significantly. In particular, it is enough to consider only one score $S_{t,s}$; the smaller $t-s$, the more significant the score should be. The different cutoffs in ~\Cref{fig:simulation_results}a are close together; therefore, all scores $S_{t,s}$ for varying $s$ are very similar. The weights ultimately reported are influenced by the randomness of the stochastic gradient descent.
For all other types of anomalies, the contrary is the case, i.e., all neighboring scores contribute significantly to the aggregated score.

Evaluating the aggregated scores on unseen test samples of the anomalous datasets, we demonstrate qualitatively (\Cref{fig:simulation_results}b) and quantitatively (\Cref{supptab:simulation_results_quantitative}) that the anomaly detection framework accurately identifies diverse types of anomalies, with F1 scores of 0.95, 0.93, 0.92, and 0.96 for drift, diffusion, noise, and spike anomalies, respectively. Furthermore, an ablation study shows that enlarging the training set improves the fidelity of the inferred conditional distributions and enhances the precision of the anomaly scores (\Cref{sec:Ablation study: decreasing the size of the training set}).

\subsection{Application of anomaly framework on irregularly observed gut microbiome dataset}\label{result:application_microbiome}

\begin{figure}[hp]
    \centering
    \includegraphics[width=\linewidth]{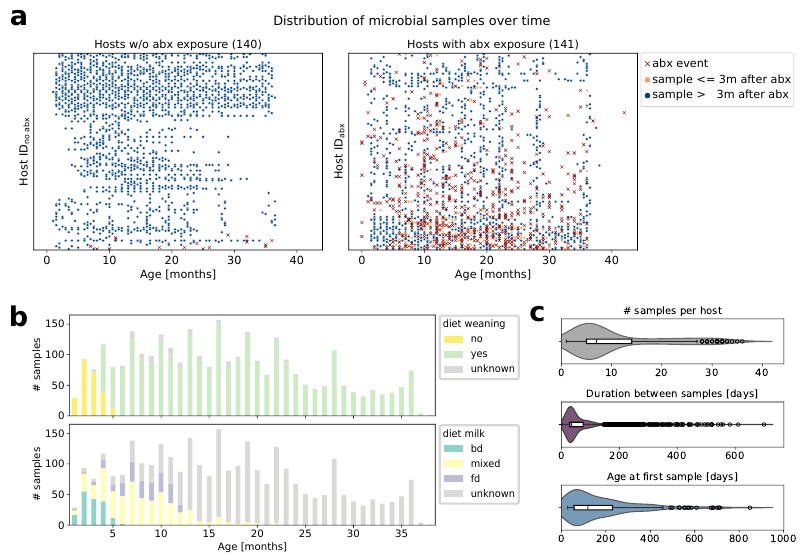}
    \caption{Description of microbiome cohort used for training and evaluation of the anomaly framework
    \textbf{(a)} Distribution of individual microbiome samples of infants with no antibiotics (abx) exposure prior to microbial sample collection (1st column) and infants with microbial samples after abx exposure (2nd column). Orange color highlights samples that were collected up to 3 months after abx exposure. The darkred crosses denote the time points of abx exposure.
    \textbf{(b)} Distribution of diet weaning and diet milk covariates in microbial samples over the infant's age.
    \textbf{(c)} Distribution of sample characteristics in the cohort.
    }
    \label{fig:dataset}
\end{figure}

\begin{figure}[hp]
    \centering
    \includegraphics[width=\linewidth]{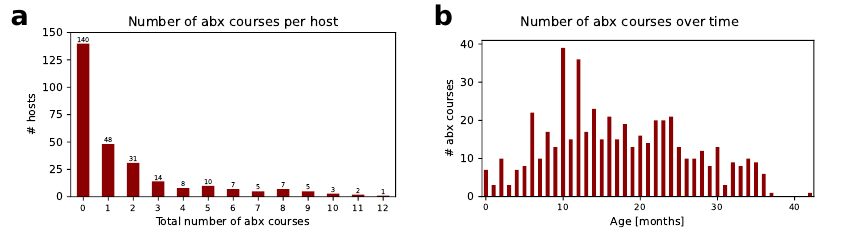}
    \caption{Description of antibiotic (abx) administrations in microbiome cohort used for evaluation of the anomaly framework.
        \textbf{(a)} Total number of abx courses per host.
        \textbf{(b)} Distribution of abx courses over age range.
    }
    \label{suppfig:abx_freq}
\end{figure}

To train and evaluate our anomaly framework on a real-world dataset, we pooled and reprocessed all datasets collected by the DIABIMMUNE Microbiome Project \cite{vatanen2019,yassour2016,vatanen2016,kostic2015} (\Cref{data_fetch,meta_proc,seq_proc}). The resulting dataset contains irregularly sampled gut microbiome profiles of 281 infants over the first 3 years of life with standardized metadata on diet, delivery mode, and antibiotic exposures (\Cref{fig:dataset}a,b; \Cref{supptab:cohort_counts}). 79.6~\% of all samples were collected in the first 2 years of life. The median sampling rate per infant was 7 samples with a median of 35 days between individual samples (\Cref{fig:dataset}c). Prior to the collected microbial samples, 140 infants had no exposure to antibiotics, and 141 infants were exposed to antibiotics, of which 66~\% had at most 3 antibiotics administrations in the observed time period (\Cref{suppfig:abx_freq}).

\begin{figure}[hp]
    \centering
    \includegraphics[width=\linewidth]{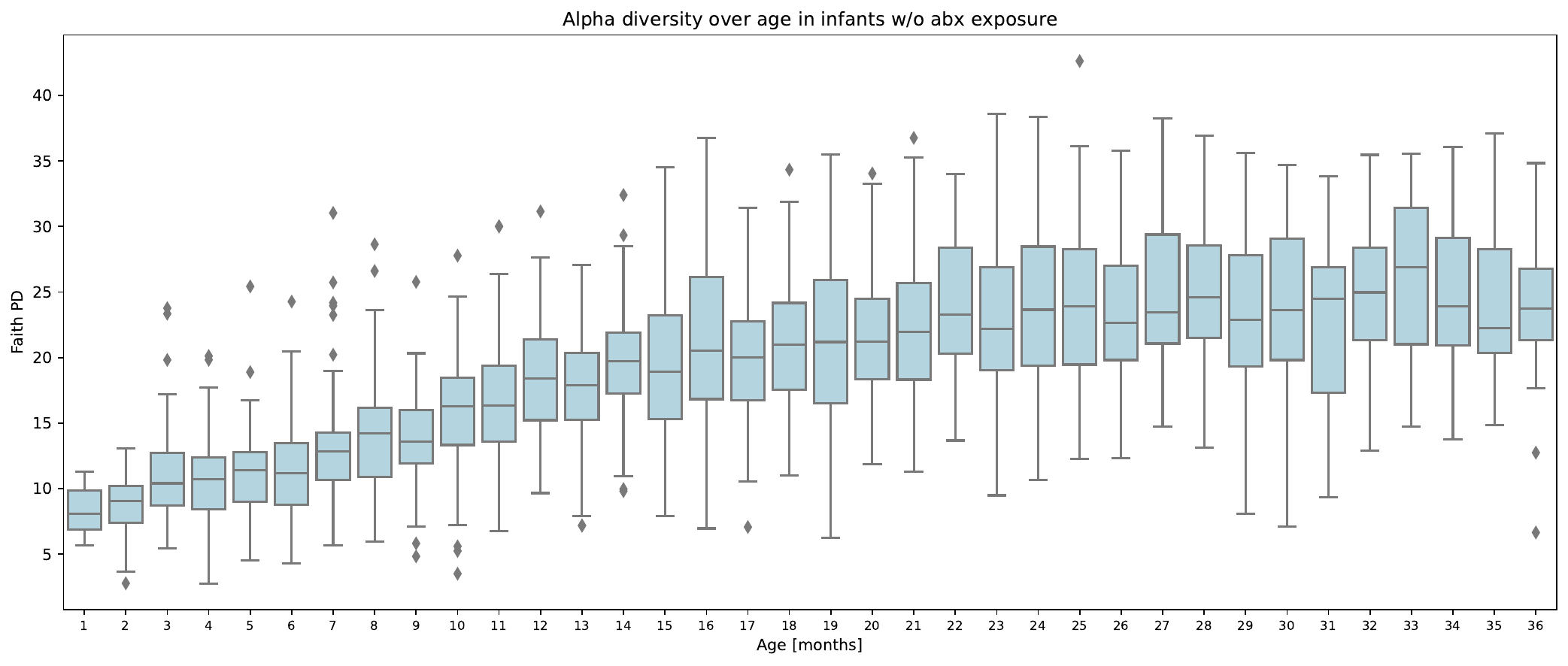}
    \caption{Distribution of alpha diversity over age among infants unexposed to antibiotics (abx). Boxplots span the 25th to 75th percentile, with the horizontal line indicating the median.}
    \label{suppfig:alpha_noabx}
\end{figure}

\begin{figure}[hp]
    \centering
    \includegraphics[width=0.8\linewidth]{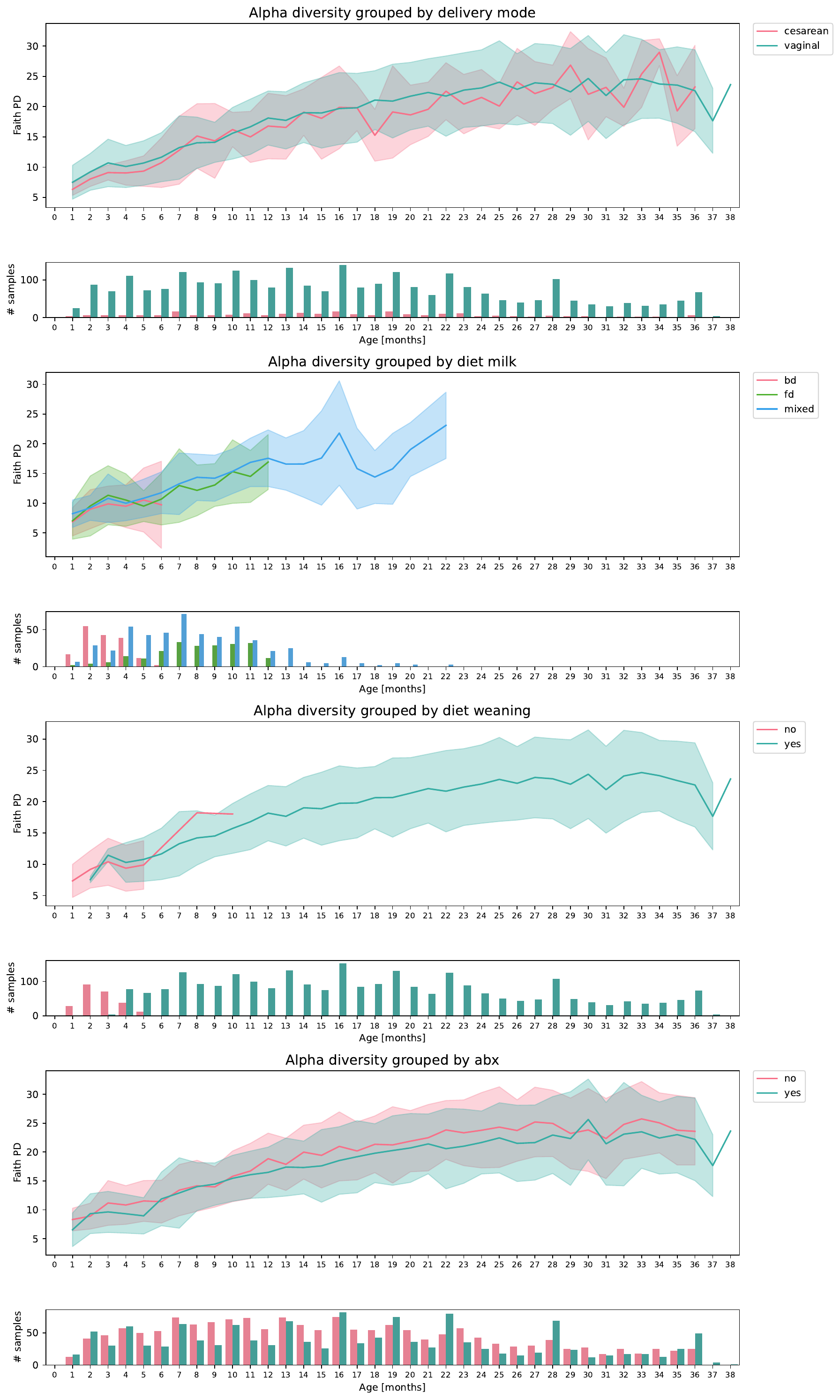}
    \caption{Temporal distribution of alpha diversity across infant age, stratified by covariates (abx = antibiotics, bd = breast milk dominant, fd = formula dominant). Upper panels show category means (lines) with standard deviations (shaded regions); standard deviations are omitted for months with a single sample per category. Lower panels display monthly sample sizes per category.}
    \label{suppfig:alpha_all_cov}
\end{figure}

As the target process $X$ for our anomaly framework, we selected the alpha diversity metric Faith's phylogenetic diversity (Faith PD) \cite{faith1992}, a summary statistic that quantifies within-sample microbial diversity by incorporating species richness and evolutionary relationships. Alpha diversity exhibits well-documented temporal patterns during infant development, showing consistent increases with age \cite{yatsunenko2012,palmer2007}, systematic variations in response to dietary changes and delivery mode \cite{bokulich2016, stewart2018, shao2019}, and decreases caused by antibiotic exposures \cite{mcdonnell2021, bokulich2016} - patterns that were also present in our dataset (\Cref{suppfig:alpha_noabx,suppfig:alpha_all_cov}). We additionally conditioned our target distribution on the infant's delivery mode and its changing dietary habits (milk diet and weaning), supplied to the model as the covariate process $C$ in \eqref{equ:PD-NJ-ODE}.

\begin{figure}[hp]
    \centering
    \includegraphics[width=1.\textwidth]{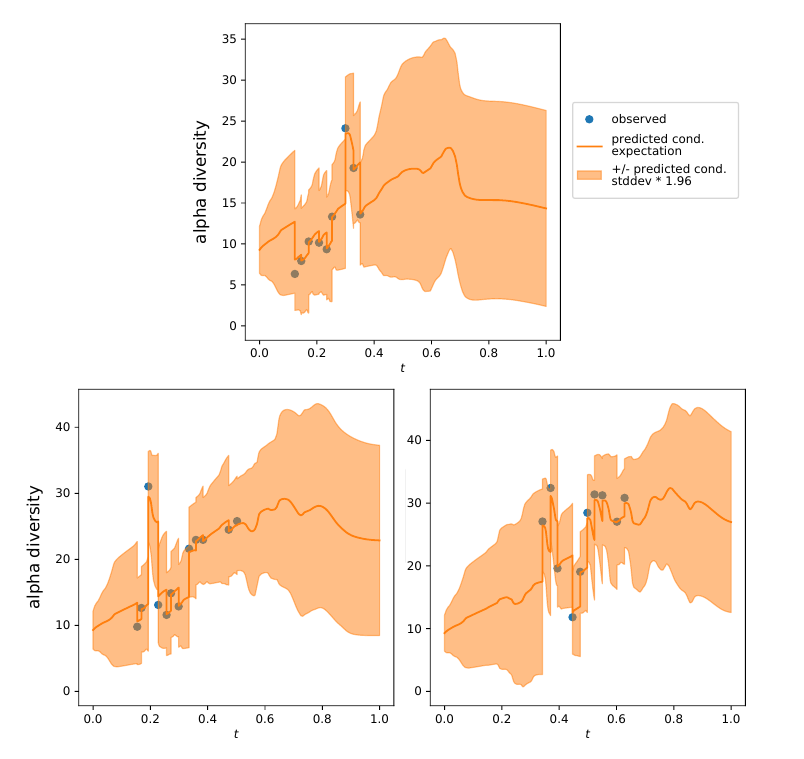}
    \caption{Examples of learned conditional expectation and standard deviation with the NJODE model trained on the microbial dataset.}
    \label{suppfig:learned_paths}
\end{figure}

\begin{figure}[hp]
    \centering
    \includegraphics[width=0.75\linewidth]{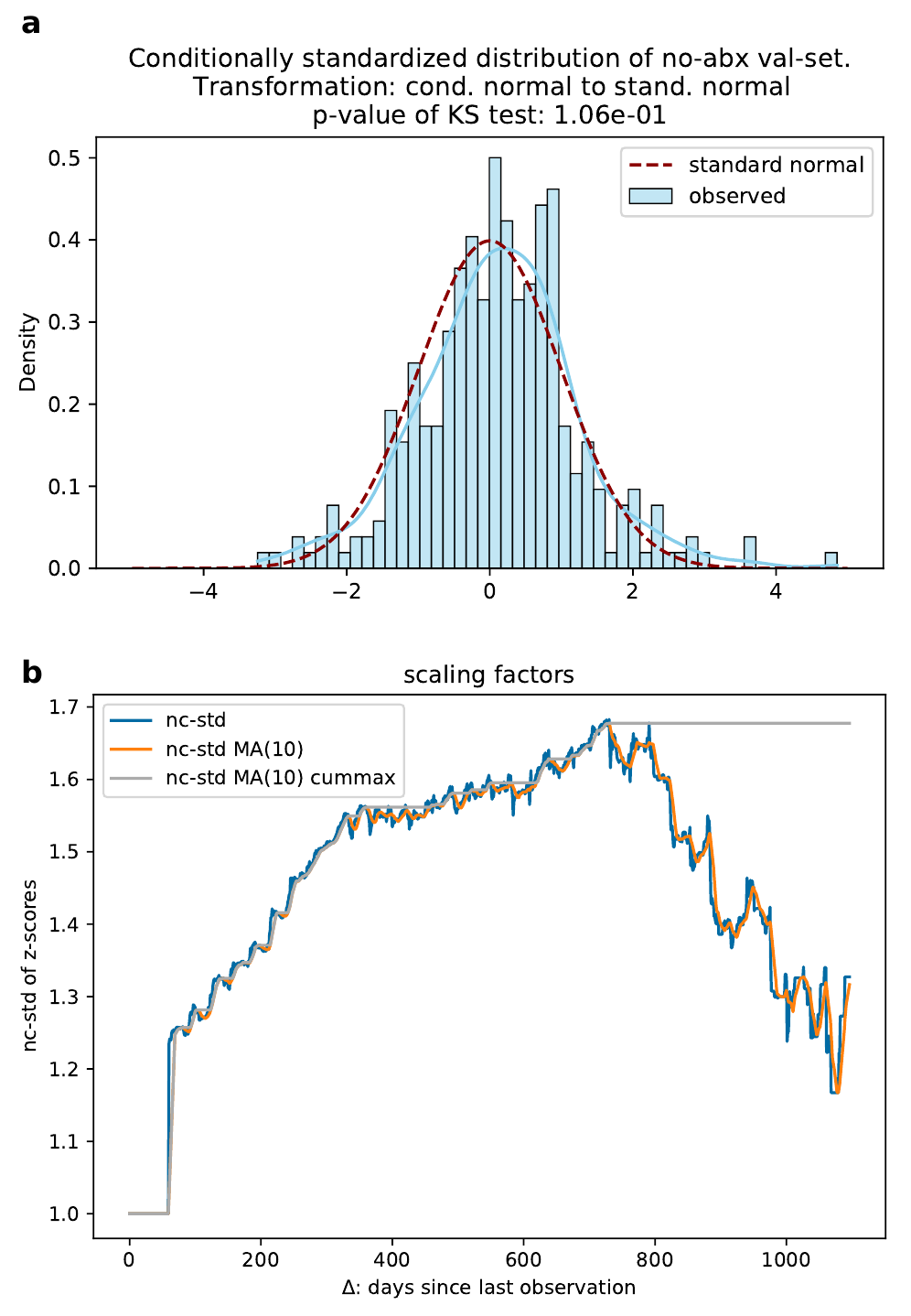}
    \caption{
        Inferred \textbf{(a)} conditional target distribution and \textbf{(b)} scaling factors on the (no-antibiotics) validation set.
        \textbf{(a)} Distribution of the conditionally standardized 1-step ahead predictions on the (no-antibiotics) validation set. For each observation $x^{(i)}$ the conditional mean $\mu^{(i)}$ and standard deviation $\sigma^{(i)}$ are predicted (based on the information up to the previous observation) and the observation (assumed to have a conditionally normal distribution) is transformed to a standard normal distribution z-score as $z^{(i)} = (x^{(i)} - \mu^{(i)})/\sigma^{(i)}$. Comparing the distribution of the z-scores of all 1-step ahead predictions (blue, with estimated density as solid line) with a standard normal probability density function (red dashed line) in a Kolmogorov-Smirnov test, does not result in a significant difference, meaning that the null-hypothesis that both distributions are the same cannot be rejected.
        \textbf{(b)} The non-centered standard deviations of z-scores (nc-std) computed in moving windows of size 120 centered around the days since last observation $\Delta$ for $\Delta \ge 60$, together with the moving average (MA) over 10 days and the cumulative maximum thereof. See \Cref{sec:Details for the computation of scaling factors} for more details.}
        \label{suppfig:inf_distribution}
\end{figure}

The antibiotics-unexposed infants were split into 80\%-20\% train-validation sets, while those exposed to antibiotics were assigned to a test set for evaluation purposes only. After training the NJODE model on the train set (\Cref{suppfig:learned_paths}, \Cref{sec:Details for the NJODE model}), we inferred the first two conditional moments on the validation set to estimate the target conditional distribution on the one-step-ahead predictions (steps 1 and 2 in  \Cref{fig:schematic}b and \Cref{sec:The anomaly framework design}), resulting in a standard normal distribution as best fit (Kolmogorov-Smirnov test was insignificant with $p~=0.106$, \Cref{suppfig:inf_distribution}a). The validation set was further used to calculate the scaling factors for the multi-step-ahead predictions (step~3 in \Cref{fig:schematic}b and \Cref{sec:Details for the computation of scaling factors}, \Cref{suppfig:inf_distribution}b). Given prior evidence that alpha diversity decreases after antibiotic exposure \cite{mcdonnell2021, bokulich2016}, we used a left-sided p-value $p$ to score observations from the train, validation, and test sets with scores $S = -\ln(p)$ (step 4 in \Cref{fig:schematic}b and \Cref{sec:The anomaly framework design}). Drawing on the analogy to spike‐type anomalies (see \Cref{sec:Microbiome data description})
in our synthetic experiments (\Cref{result:simulate}), we interpreted anomaly scores as direct indicators of antibiotic‐induced perturbations.

As verification that the scaling factors adjust the predictions for the epistemic uncertainty of multi-step predictions, we generated forecasts at selected random temporal cut-off points within the validation set and compared the scores for increasing $\Delta$ to the scores prior to the cut-off. Our results indicate that the model accurately predicts for at least 12 months post cut-off, with inferred distributions matching the observed data (\Cref{suppfig:time_horizon}).

\subsection{Description of anomalies from delineated antibiotic exposures}\label{result:describe1}

\begin{figure}[hp]
    \centering
    \includegraphics[width=\linewidth]{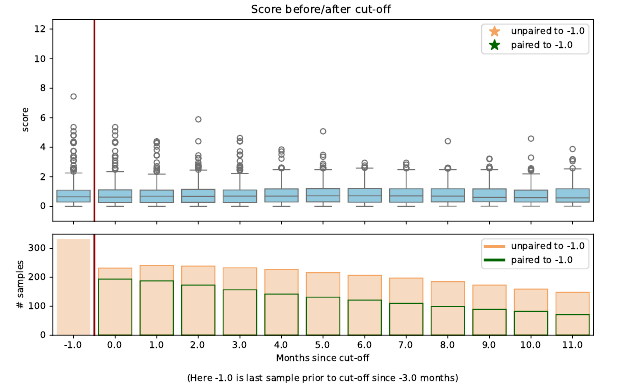}
    \caption{Distributions of anomaly scores prior and after selected cut-offs in validation set depicting the anomaly framework's reliable multi-step ahead prediction horizon. Red vertical lines mark cut-off timings. Stars denote the statistical significance comparison of anomaly scores pre- and post-cut-off (* $p < 0.1$, ** $p < 0.05$), where yellow stars represent Mann-Whitney U-tests and green stars represent Wilcoxon tests. The lower plots display the number of samples available within each monthly time bin, with positive x-axis values representing intervals that include the left boundary (e.g., $x=0$ corresponds to $[0,1)$) and $x=-1$ representing the last sample observed in the 3 months prior to cut-off.
    }
    \label{suppfig:time_horizon}
\end{figure}

\begin{figure}[hp]
    \centering
    \includegraphics[width=\linewidth]{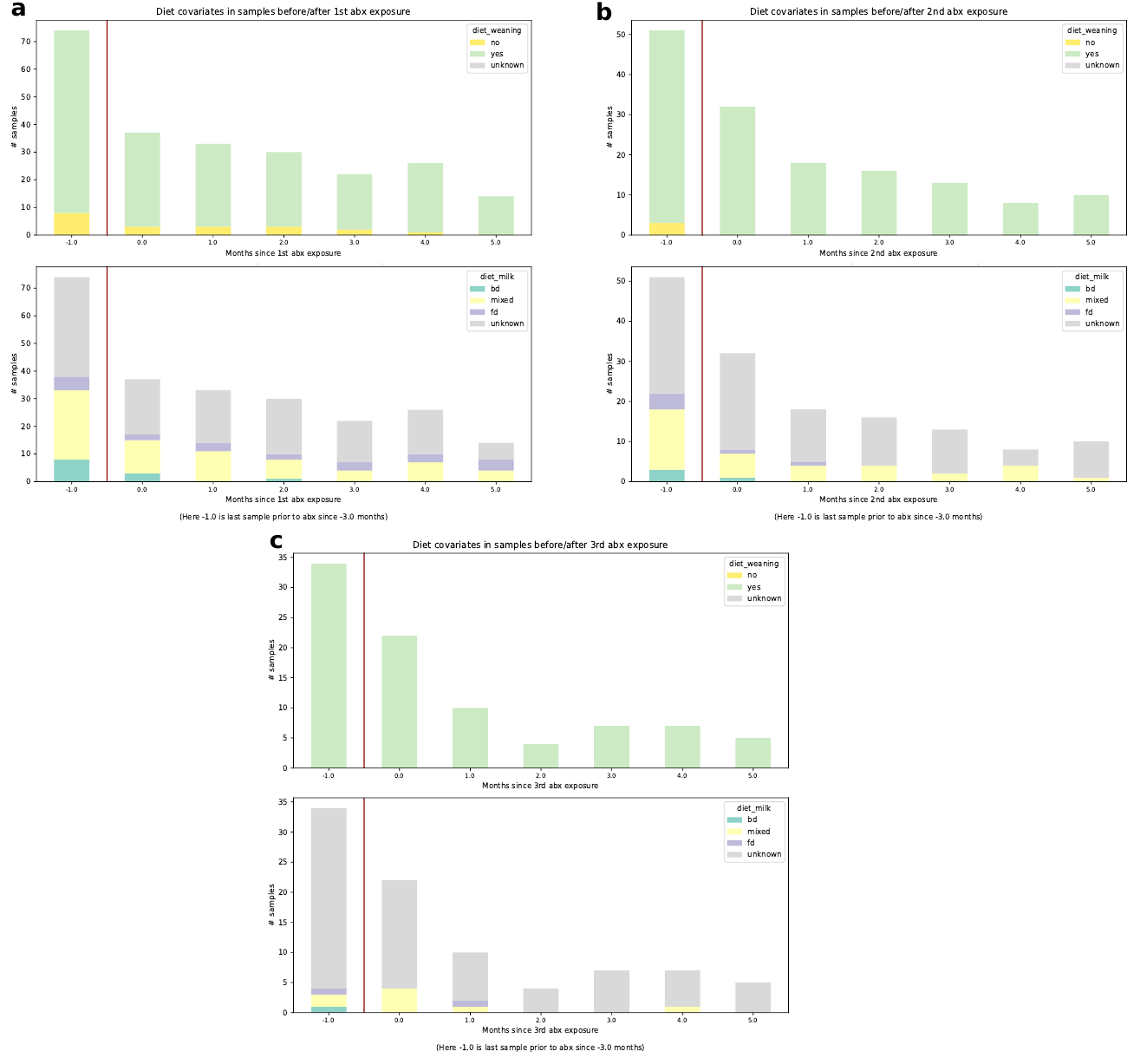}
    \caption{Prevalence of dietary habits before and after \textbf{(a)}~first, \textbf{(b)}~second and \textbf{(c)}~third antibiotic exposure (bd = breast milk dominant, fd = formula dominant). Red vertical lines indicate the timing of each antibiotic exposure. Positive x-axis values represent intervals that include the left boundary (e.g., $x=0$ corresponds to $[0,1)$) and $x=-1$ represents the last sample observed in the 3~months prior to antibiotic exposure.}
    \label{suppfig:diet_covariate_post_abx}
\end{figure}

To investigate the duration and magnitude of the anomalies that follow individual antibiotic exposures, we scored all observations of the antibiotic-exposed infants in the test set based on the inferred distribution derived from unexposed infants (\Cref{result:application_microbiome}). We limited our analysis to observations within the first two years of life due to reduced sampling frequency in later age ranges (\Cref{fig:dataset}a). We focused on effects lasting up to six months post-exposure with enough samples available, even though longer predictions could reliably be used (\Cref{suppfig:time_horizon}).
To delineate the effects of subsequent exposures, we evaluated the anomalies after the first, second, and third antibiotic administration by using the time point of administration as the cut-off for multi-step-ahead predictions. After the cut-offs, the predicted distributions were dynamically conditioned on the covariate process $C$, i.e., the delivery mode and changes in diet (for prevalence of dietary habits pre- and post-antibiotics exposure see \Cref{suppfig:diet_covariate_post_abx}). Individual anomaly score trajectories of selected infants reveal an increase following each of the first three antibiotic exposures (\Cref{fig:score}a).

We observed that anomalies arising from the second antibiotic exposure were most pronounced and persisted longer than those following the first or third exposures (\Cref{fig:score}b). After the second exposure, scores remained significantly increased for up to 4 months compared to pre-exposure levels. Both the first and second antibiotic exposures contained enough samples post exposure and displayed similar characteristics, differing primarily in their timing, as the first exposure occurred earlier in the infants' development (\Cref{fig:score}b, \Cref{supptab:first_second_abx_char}).

We compared these dynamically inferred anomaly profiles with a static alpha diversity matching technique, in which alpha diversity differences before and after antibiotic exposures were computed by subtracting the mean alpha diversity of matched unexposed samples from that of antibiotic-exposed samples. Matching was performed via monthly age bins, delivery mode, and dietary status (milk feeding and weaning). Both approaches yielded similar results (\Cref{fig:score}b,c). However, the dynamic framework offers several advantages: (i) it dynamically captures temporal trajectories beyond coarse monthly bins, (ii) it incorporates complex relationships between alpha diversity, delivery mode, and dietary changes rather than relying on pre-specified confounders, and (iii) it provides a foundation for further extensions (for instance, modeling additional target features), whereas the static matching approach is constrained to a single target.

\begin{figure}[hp]
    \centering
    \includegraphics[width=0.7\linewidth]{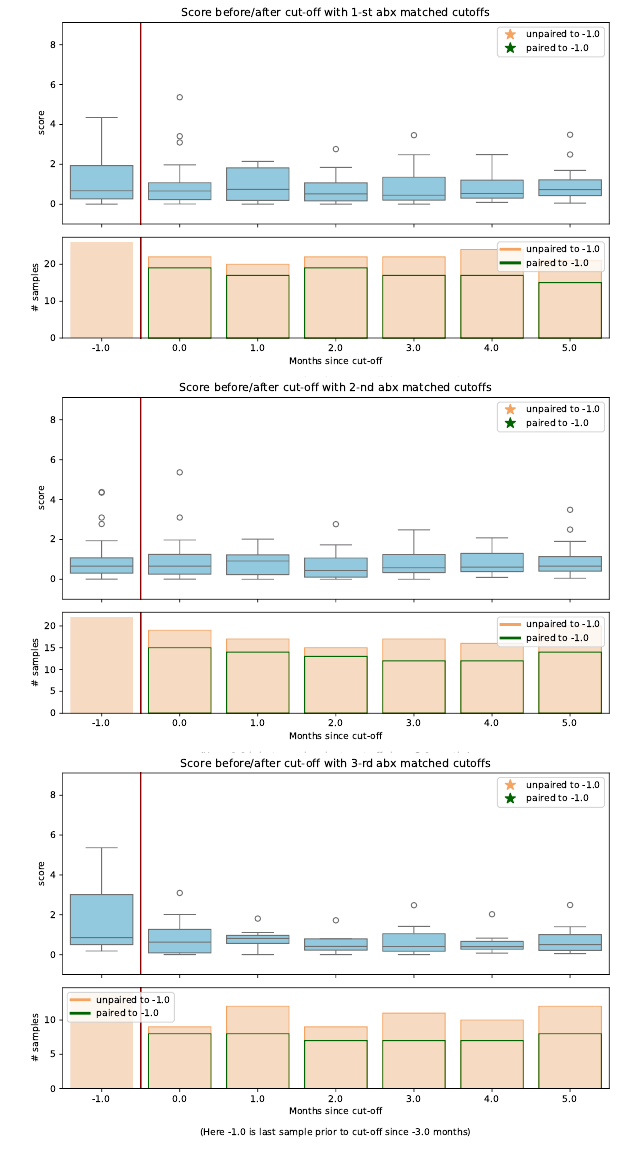}
    \caption{Benchmark analysis of anomaly scores in the validation set (which does not contain any abx exposures), using temporal cut-offs (vertical red lines) corresponding to antibiotic exposure timepoints in the test set. Stars denote the statistical significance of the difference in the metric distribution post cut-off compared to values preceding the cut-off (* $p < 0.1$, ** $p < 0.05$), where yellow stars represent Mann-Whitney U-tests and green stars represent Wilcoxon tests. The lower plots display the number of samples available within each monthly time bin, with positive x-axis values representing intervals that include the left boundary (e.g., $x=0$ corresponds to $[0,1)$) and $x=-1$ representing the last sample observed in the 3~months prior to cut-off.}
    \label{suppfig:baseline}
\end{figure}

\begin{figure}[hp]
    \centering
    \includegraphics[width=\linewidth]{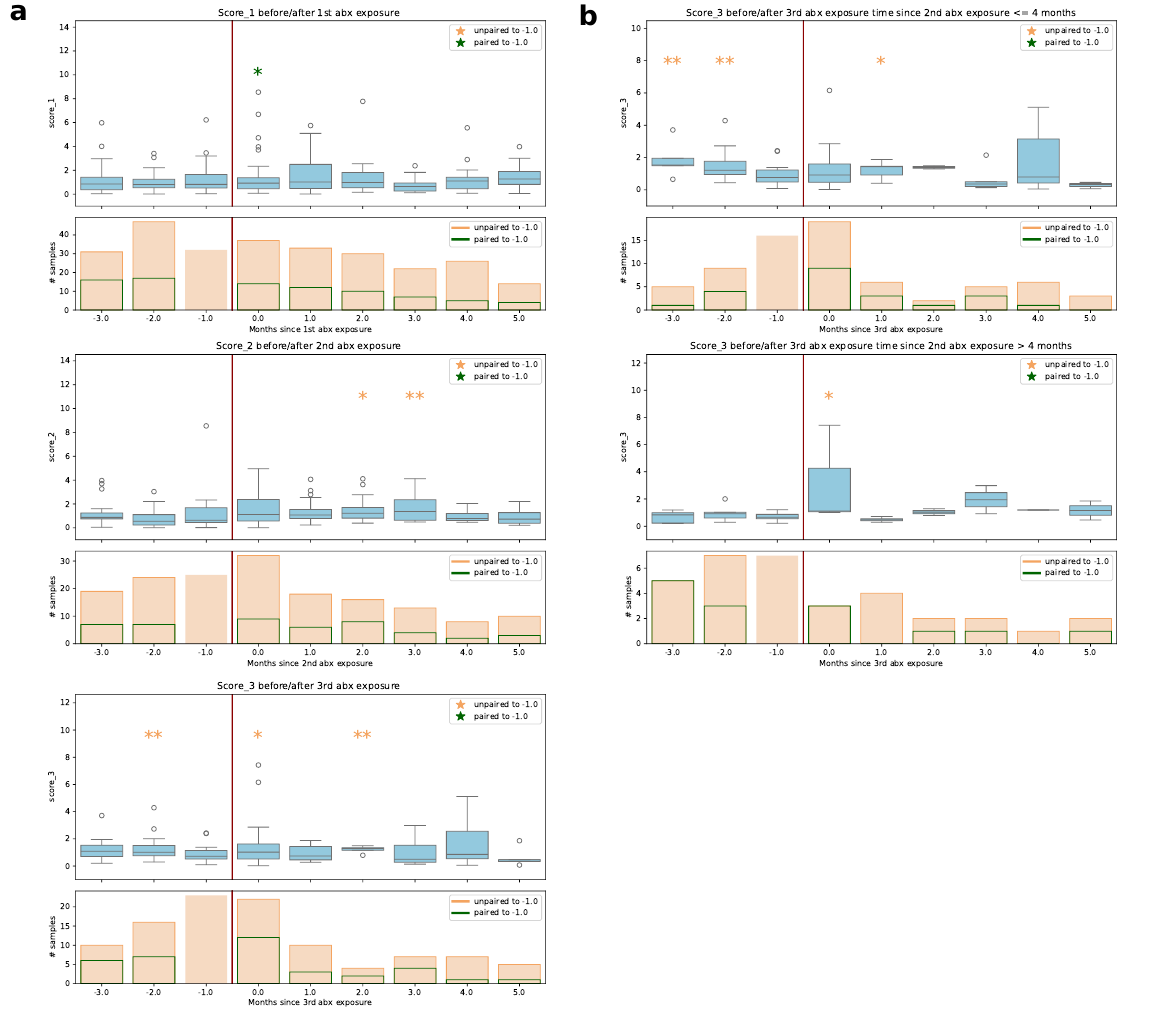}
    \caption{
    Distributions of anomaly scores split by monthly time bins prior and after \textbf{(a)}~1st, 2nd and 3rd antibiotics exposures and \textbf{(b)}~3rd antibiotics exposure split by time since 2nd exposure. Red vertical lines indicate the timing of each antibiotic exposure. Stars denote the statistical significance of the difference in the metric post-exposure compared to values preceding exposure (* $p < 0.1$, ** $p < 0.05$), where yellow stars represent Mann-Whitney U-tests and green stars represent Wilcoxon tests. The lower plots display the number of samples available within each monthly time bin, with positive x-axis values representing intervals that include the left boundary (e.g., $x=0$ corresponds to $[0,1)$).
    }
    \label{suppfig:priorexposure}
\end{figure}

As an additional benchmark for the dynamically inferred anomaly profiles, we extracted matching time points from the validation set corresponding to antibiotic exposures in the test set and found no significant differences in anomaly score distributions (\Cref{suppfig:baseline}). By segmenting the three-month age bins preceding antibiotic exposures, we observed stable anomaly score distributions before the first and second antibiotic exposures (\Cref{suppfig:priorexposure}a). In contrast, an increase in anomaly scores was detected two months prior to the third antibiotic exposure, which is attributable to lingering effects from closely preceding second antibiotic exposures (\Cref{suppfig:priorexposure}b).

These findings demonstrate that our dynamic anomaly scoring approach not only robustly quantifies the longitudinal impact of individual consecutive antibiotic exposures on the infant microbiome but also flexibly accommodates complex factors such as delivery mode and dietary transitions, providing a versatile framework for extended applications in microbiome research and beyond.

\begin{figure}[p]
    \centering
    \includegraphics[width=\linewidth]{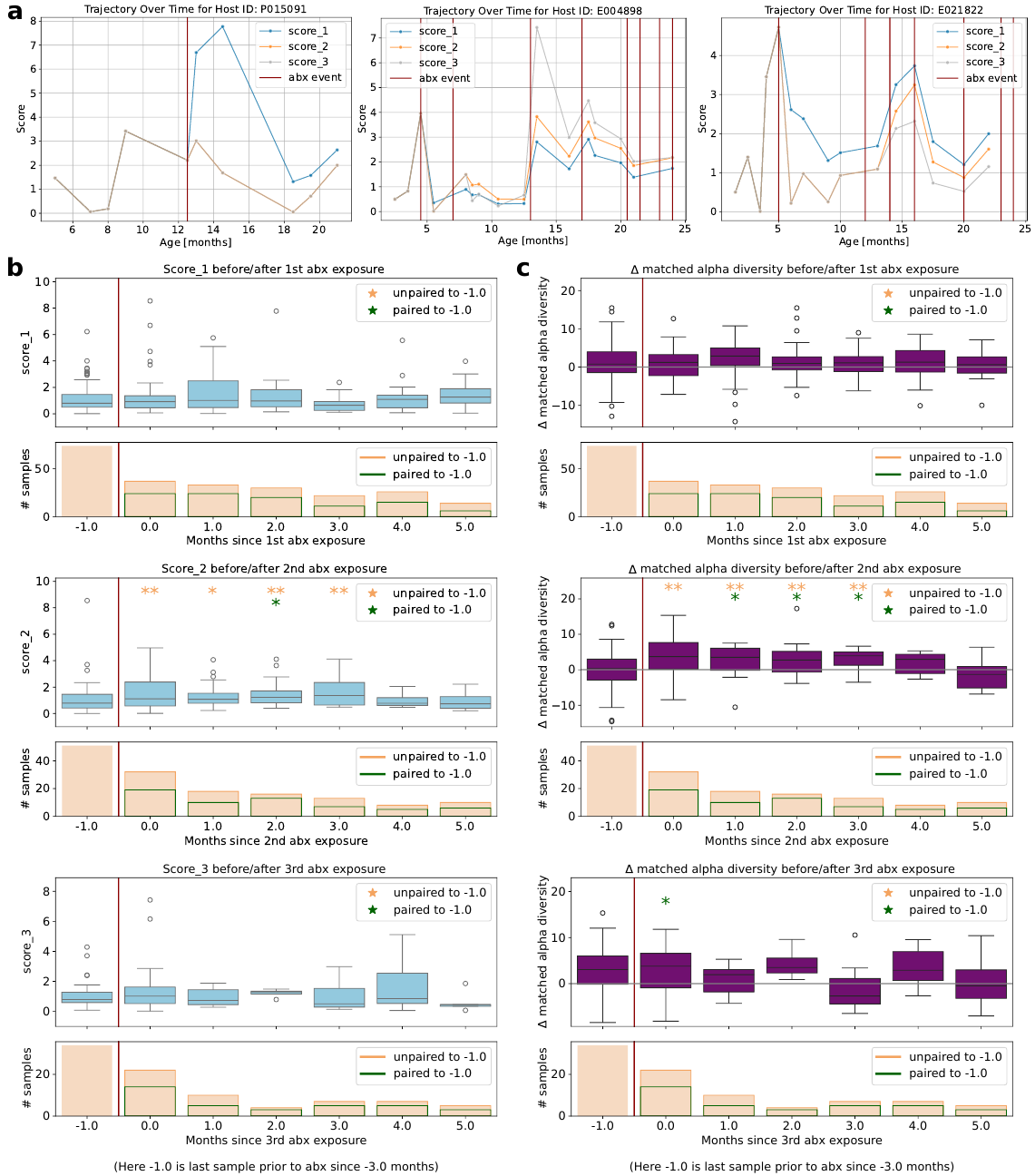}
    \end{figure}
    
    \begin{figure}[t!]
    \caption{Insights on anomaly dynamics from individual antibiotics exposures. \textbf{(a)} Individual host trajectories of multi-step ahead scores after the first (score\_1), second (score\_2) and third (score\_3) antibiotic exposures. Because antibiotic exposure times are recorded at half‑month resolution, the first exposure appears concurrent with an alpha diversity observation (see \Cref{sec:DataResolutionProblem}); however, the exposure may have occurred slightly after the observation in the first plot or slightly before it in the second.
    \textbf{(b + c)} Distributions of metrics, \textbf{(b)} anomaly scores and \textbf{(c)} alpha diversity differences, prior and after antibiotics exposures. Red vertical lines indicate the timing of each antibiotic exposure. Stars denote the statistical significance of the difference in the metric post-exposure compared to values preceding exposure (* $p < 0.1$, ** $p < 0.05$), where yellow stars represent Mann-Whitney U-tests and green stars represent Wilcoxon tests. The lower plots display the number of samples available within each monthly time bin, with positive x-axis values representing intervals that include the left boundary (e.g., $x=0$ corresponds to $[0,1)$) and $x=-1$ representing the last sample observed in the 3~months prior to antibiotic exposure.
    In \textbf{(c)}, the difference in alpha diversity was calculated as the mean alpha diversity of matched unexposed samples minus that of antibiotic-exposed samples. Samples were matched based on monthly age bin, delivery mode, and dietary status (milk feeding and weaning).
    }
    \label{fig:score}
\end{figure}

\subsection{Dynamics of first and second antibiotic exposure depend on administration duration and time of life}\label{result:describe2}

\begin{figure}[p]
    \centering
    \includegraphics[width=0.95\linewidth]{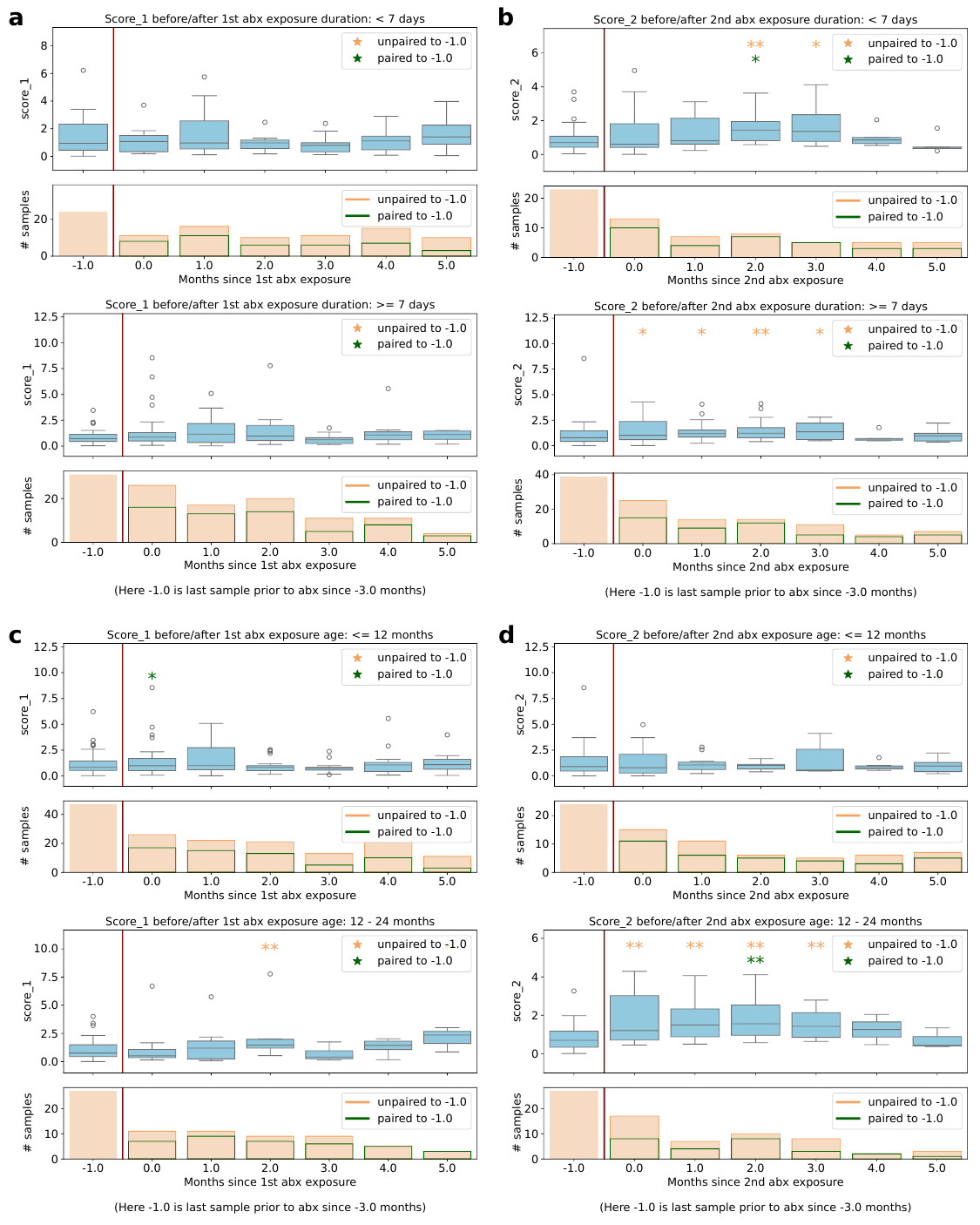}
    \caption{Distributions of anomaly scores prior and after individual antibiotics exposures split by \textbf{(a,b)} duration of antibiotics exposure and by \textbf{(c,d)} age of infant at exposure. Red vertical lines indicate the timing of the \textbf{(a,c)} 1st and \textbf{(b,d)} 2nd antibiotic exposures. Stars denote the statistical significance of the difference in the metric post-exposure compared to values preceding exposure (* $p < 0.1$, ** $p < 0.05$), where yellow stars represent Mann-Whitney U-tests and green stars represent Wilcoxon tests. The lower plots display the number of samples available within each monthly time bin, with positive x-axis values representing intervals that include the left boundary (e.g., $x=0$ corresponds to $[0,1)$) and $x=-1$ representing the last sample observed in the 3~months prior to antibiotic exposure.}
    \label{fig:scoresplit}
\end{figure}

\begin{figure}[p]
    \centering
    \includegraphics[width=0.95\linewidth]{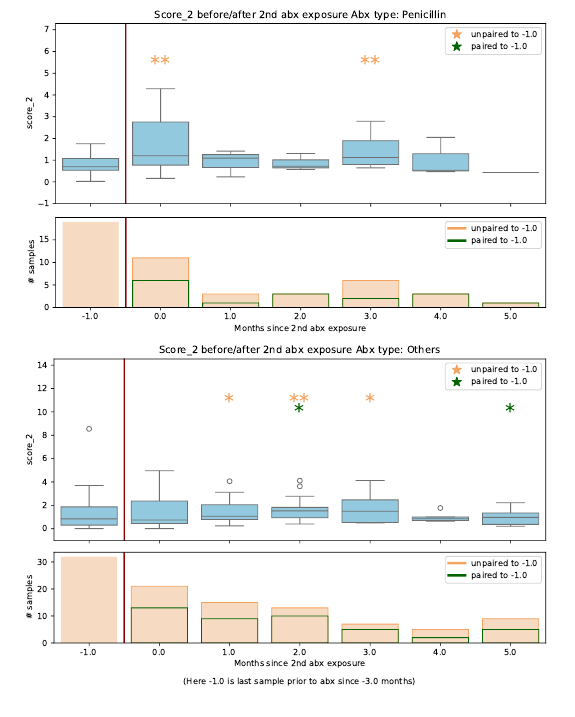}
    \caption{Distributions of anomaly scores prior and after 2nd antibiotics exposure split by type of antibiotic that was prescribed. Red vertical lines indicate the timing of the 2nd antibiotic exposures. Stars denote the statistical significance of the difference in the metric post-exposure compared to values preceding exposure (* $p < 0.1$, ** $p < 0.05$), where yellow stars represent Mann-Whitney U-tests and green stars represent Wilcoxon tests. The lower plots display the number of samples available within each monthly time bin, with positive x-axis values representing intervals that include the left boundary (e.g., $x=0$ corresponds to $[0,1)$) and $x=-1$ representing the last sample observed in the 3~months prior to antibiotic exposure.}
    \label{suppfig:penicillin2nd}
\end{figure}

\begin{figure}[p]
    \centering
    \includegraphics[width=0.95\linewidth]{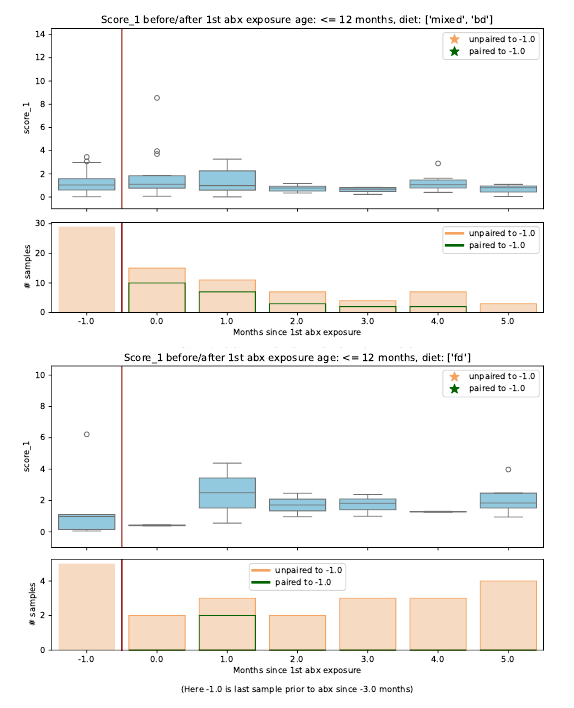}
    \caption{Distributions of anomaly scores prior and after 1st antibiotics exposure in first year of an infant's life split by milk diet (bd = breast milk dominant, fd = formula dominant). Red vertical lines indicate the timing of the 1st antibiotic exposures. Stars denote the statistical significance of the difference in the metric post-exposure compared to values preceding exposure (* $p < 0.1$, ** $p < 0.05$), where yellow stars represent Mann-Whitney U-tests and green stars represent Wilcoxon tests. The lower plots display the number of samples available within each monthly time bin, with positive x-axis values representing intervals that include the left boundary (e.g., $x=0$ corresponds to $[0,1)$) and $x=-1$ representing the last sample observed in the 3~months prior to antibiotic exposure.}
    \label{suppfig:scoresplit_age_diet}
\end{figure}

We further investigated how antibiotic exposure characteristics affect the observed anomaly patterns. Overall prolonged antibiotic courses ($\geq 7$ days) yielded pronounced, sustained anomalies following the second exposure (\Cref{fig:scoresplit}b) but not following the first exposure (\Cref{fig:scoresplit}a). After the second exposure, Penicillin, the most commonly administered antibiotic, had a more short-term anomalous effect than other administered antibiotic types (\Cref{suppfig:penicillin2nd}, consistent with \cite{wurm2024}). Additionally, the timing of administration influenced anomaly severity: exposures occurring in the second year of life displayed more pronounced anomalies than those within the first year (\Cref{fig:scoresplit}c,d). In the first year, only infants with no breast milk exposure showed increased anomaly scores post-exposure (\Cref{suppfig:scoresplit_age_diet}), suggesting that breastmilk feeding may help "rescue" age-normative microbiome development following antibiotic exposure early in life.

\begin{figure}[p]
    \centering
    \includegraphics[width=\linewidth]{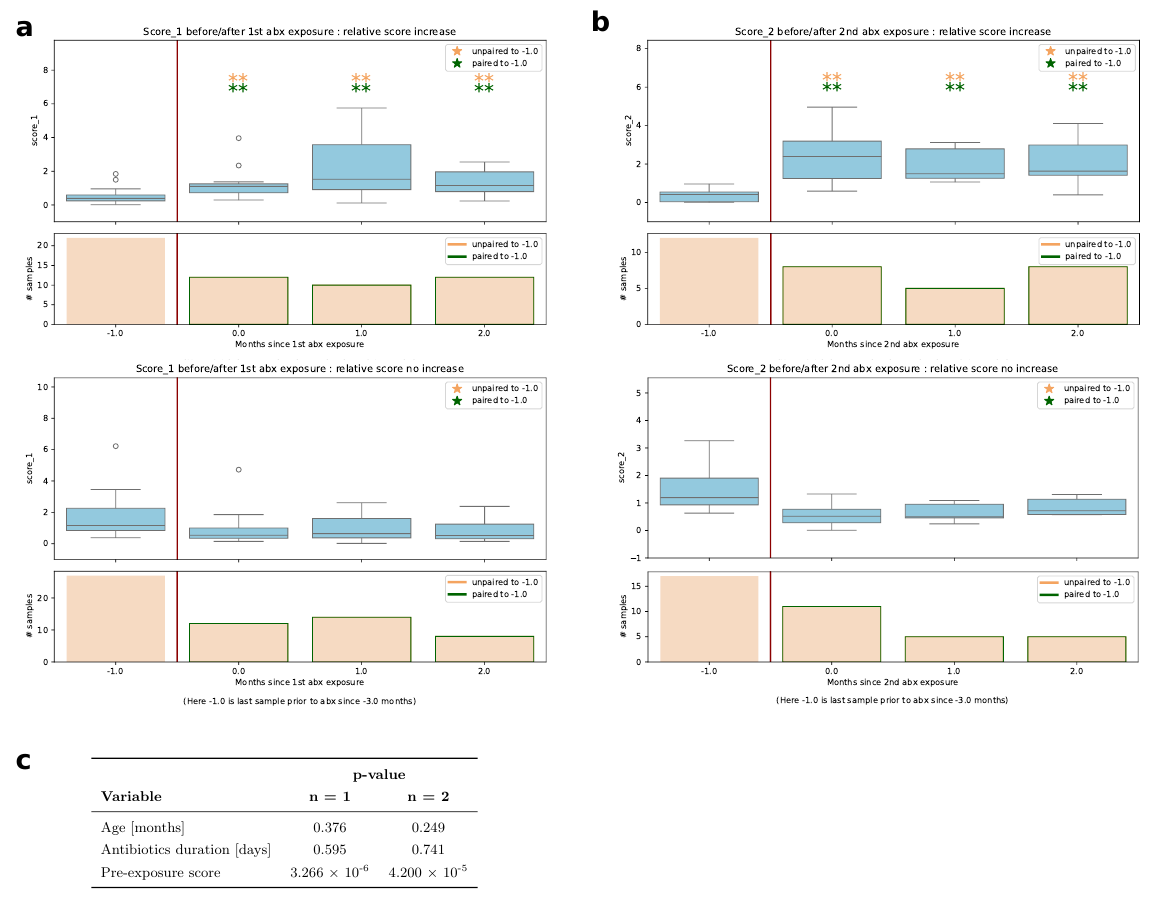}
    \caption{Comparative analysis of antibiotic exposure characteristics and anomaly scores stratified by post-antibiotic response.
        \textbf{(a,b)} Distribution of anomaly scores after \textbf{(a)} first and \textbf{(b)} second antibiotic exposures, with cohorts divided based on presence or absence of notable $\geq 2$\nobreakdash-fold score increases post-exposure. This analysis is restricted to infants with at least one microbial sample before and after exposure.
        \textbf{(c)}~Statistical comparison of antibiotic exposure characteristics and anomaly score distributions between response groups using Mann-Whitney U tests, shown separately for first (n=1) and second (n=2) antibiotic exposures.}
    \label{suppfig:indiv_score_analysis}
\end{figure}

When comparing antibiotic exposure characteristics of infants who exhibited $\geq 2$\nobreakdash-fold increases in their post-antibiotic anomaly scores with those who did not, no significant differences emerged (\Cref{suppfig:indiv_score_analysis}), which may be attributable to sample-size constraints imposed by the requirement for at least one microbial sample before and after each exposure. Nevertheless, one exception was observed: infants without notable post-exposure increases exhibited a significantly elevated pre-exposure score distribution (\Cref{suppfig:indiv_score_analysis}c). This highlights a limitation of this analysis to describe microbial anomalies in infants where the microbiome is already perturbed prior to antibiotic exposure (\Cref{appendix:causality}) and a limitation in the dataset's resolution (\Cref{sec:DataResolutionProblem}). A larger dataset containing more detailed, temporal pre-antibiotic exposure information would allow us to exploit the full potential of our anomaly detection framework in untangling the multifaceted disruptions impacting the infant gut microbiome.

\subsection{Predictive models based on anomaly scores outperform diversity-based baselines}\label{result:predict}
We further assessed whether the model's anomaly scores can be used to predict antibiotic exposure events using one-step-ahead score predictions (\Cref{sec:predictive_setup}). Three anomaly score-based predictions were compared against their respective baseline and random predictions (\Cref{tab:prediction_table}). In the score-based (S) predictions, the anomaly scores were used to predict antibiotic exposure events, whereas in the baseline (B) predictions, the difference of observed to matched alpha diversity or static metadata values were used as features. Antibiotic events were predicted via (1) an absolute threshold on the feature value, (2) a relative threshold for the increase of the respective feature between consecutive samples within three months, and (3) a Random Forest classifier \cite{breiman2001}.

All score-based predictions outperformed the baseline and random predictions (\Cref{tab:prediction_metrics}), highlighting that our derived anomaly scores managed to capture dynamics from antibiotics exposures that were not reflected in basic features used in the baselines, namely alpha diversity and basic metadata features. These results demonstrate the robustness and potential clinical utility of the model's anomaly scores as sensitive indicators of perturbation.

%%===============================%%
%% Discussion %%

\section{Discussion}\label{discussion}
We present a forecasting-based anomaly detection framework that leverages NJODEs to learn the dynamics of irregularly and sparsely sampled time-series. First validated on synthetic data, where it accurately distinguishes changes in drift, diffusion, additive noise, and transient spikes, the framework was next applied to reveal the temporal anomalies induced by antibiotic exposure in the infant gut microbiome. While fully adjusting for complex confounders, such as birth mode and diet transitions, we learned a prediction model of the healthy microbial dynamics in infants---represented by alpha diversity dynamics. The learned conditional distribution was then used to derive anomaly scores of observations post-antibiotic exposure, allowing us to delineate the duration and magnitude of antibiotic-induced perturbations. 

Anomalies exhibited prolonged persistence following second antibiotic courses, extended-duration regimens, and exposures administered during the second year of life. The observed rapid normalization of the gut microbiome following antibiotic exposures in the first year of life, when most infants receive their first administration, may be attributable to breast milk exposure \cite{brockway2024, dai2023}. Breast milk contains oligosaccharides and bioactive compounds that promote beneficial microbial growth and resilience \cite{zhernakova2025}, which may explain the enhanced recovery capacity during this critical developmental window. Additionally, given the lower baseline diversity in early life \cite{yatsunenko2012}, antibiotic-induced reductions are correspondingly smaller in absolute magnitude, thereby complicating their detection as anomalies. Despite apparent rapid recovery of alpha diversity following the first antibiotic exposure, the effects on the gut microbiome may be longer-lasting (e.g., \cite{azad2014, hviid2011, kilkkinen2006, korpela2017}) and more wide-ranging, affecting individual microbial entities, the gut resistome, and its functional potential in ways not captured by diversity metrics of amplicon sequences alone \cite{gasparrini2019}.

We further showed that the inferred anomaly scores can be used to accurately detect antibiotic exposures, outperforming diversity-based baseline predictions. The superior ability of anomaly score-based models to identify individual antibiotic exposures, relative to diversity-based baselines (\Cref{tab:prediction_metrics}), indicates that static, binned adjustment (e.g., age-, diet-, and delivery mode-matched comparisons) does not capture the full temporal structure of antibiotic perturbations. While coarse, binned alpha diversity trends appear similar at the aggregate level (\Cref{fig:score}b,c), probabilistic forecasting-derived anomaly scores seem to retain individualized trajectory context, yielding additional discriminative signal for exposure detection.

Our anomaly detection framework is highly versatile, also in a data-limited setting with scarce and irregular observations, as is typical in human microbiome and many clinical intervention studies. 
While our microbiome application has several limitations — univariate focus on alpha diversity, scarce sampling rate, moderate sample size, and insufficient pre-antibiotic perturbation data — the anomaly framework is not restricted to microbiome data and can be applied on different multivariate time-series, also with more complex conditional distributions beyond the Gaussian case studied here. Given a larger microbiome cohort with better annotated and denser temporal sampling rates, it could also incorporate multivariate microbiome feature dynamics (e.g., taxonomic abundances, functional gene profiles, host biomarkers). This could reveal further details about the impact of antibiotics on the microbial temporal dynamics. Ultimately, the ability to describe and predict individual microbiome anomalies in real time holds promise for personalized monitoring of gut microbiome health and holds translational potential: by quantifying anomalies, this approach could inform and optimize clinical treatment regimens, e.g., to detect and minimize perturbations to the gut microbiome, or to time specific treatments based on anomaly scoring e.g., to time drug delivery according to microbiota perturbations that precede infection or other adverse events.

\begin{table}[hp]
    \centering
    \renewcommand{\arraystretch}{1.5}

    \caption{Overview of prediction set-ups \textbf{(a)} and their performance metrics \textbf{(b)}\label{tab:prediction_table}}

    \begin{subtable}{\textwidth}
        \centering
        \begin{tabular}{p{0.2\textwidth}p{0.4\textwidth}p{0.4\textwidth}}
            \hline
                                                                                                                        & \textbf{Score-based (S)}                                                                                & \textbf{Baseline (B)} \\[6pt]
            \hline
            \textbf{(1)} Absolute                                                                                       &
            Quantile-based ($q$) score threshold inferred from no-antibiotics validation set:                           &
            Absolute threshold ($threshold$) for observed difference to matched diversity:                                                                                                                                                                \\[6pt]
                                                                                                                        & $\text{score} > \text{threshold}(q)$                                                                    &
            $\text{observed diff} > threshold$                                                                                                                                                                                                            \\[6pt]
            \hline
            \textbf{(2)} Relative                                                                                       &
            Relative change in score from former to next sample within 3 months larger than defined value ($rel\_inc$): &
            Relative change in difference to matched diversity from former to next sample within 3 months larger than defined value ($rel\_inc$):                                                                                                         \\[6pt]
                                                                                                                        & $\frac{\text{score}(t)}{\text{score}(t-1)} \geq rel\_inc$ \newline for $t - (t-1) \leq 3\text{ months}$ &
            $\frac{\text{obs. diff}(t)}{\text{obs. diff}(t-1)} \geq rel\_inc$ \newline for $t - (t-1) \leq 3\text{ months}$                                                                                                                               \\[6pt]
            \hline
            \textbf{(3)} Random Forest                                                                                  &
            Classifier trained to predict exposure based on score only:                                                 &
            Classifier trained to predict exposure based on age, diet, delivery mode and diversity from same time point:                                                                                                                                  \\[6pt]
                                                                                                                        & $\text{RF}(\text{score}) \geq 0.5$                                                                      &
            $\text{RF}(\text{static features}) \geq 0.5$                                                                                                                                                                                                  \\[6pt]
            \hline
        \end{tabular}
        \caption{Description of score-based and baseline prediction set-ups. In (1) and (2) the absolute and relative threshold values were treated as hyperparameters and only the best performing values according to the macro-averaged F1-score was selected. In (3) classifiers were trained with default scikit-learn \cite{pedregosa2011} hyperparameters on 70\% of host-stratified samples.}
        \label{tab:prediction_description}
    \end{subtable}

    \vspace{4mm}

\begin{subtable}{\textwidth}
\centering
\begin{tabular}{p{0.2\textwidth}p{0.2\textwidth}p{0.2\textwidth}p{0.3\textwidth}}
\hline
        \textbf{Set-up}             & \textbf{Macro-avg.\ F1} & \textbf{MCC} & \textbf{Best Hyperparameters} \\
        \hline
        \textbf{S1}      & 0.542                  & 0.118        & $q = 0.68$            \\
        \hline
        \textbf{S2}      & 0.504                  & 0.018        & $rel\_inc = 3.0$      \\
        \hline
        \textbf{S3}            & 0.501                  & 0.002        & -                    \\
        \hline
        \textbf{R} class proportion & 0.500                  & 0.000        & -           \\
        \hline
        \textbf{B1}      & 0.496                  & 0.018        & $threshold = 7.0$     \\
        \hline
        \textbf{B3} & 0.494                  & 0.018        & -                    \\
        \hline
        \textbf{B2}      & 0.472                  & -0.051       & $rel\_inc = 1.4$      \\
        \hline
        \textbf{R} uniform          & 0.458                  & 0.001        & -               \\
        \hline
        \textbf{R} all negative     & 0.438                  & 0.000        & -         \\
        \hline
        \end{tabular}
    \caption{Performance metrics and best hyperparameters of score-based (S), baseline (B), and random (R) predictions. Predictive performance is evaluated with the macro-averaged F1-score (sorted decreasing) and the Matthews correlation coefficient (MCC), accounting for the imbalanced class ratios and the primary interest in predicting the positive samples correctly. Three random guessing predictions (R) are included for comparison: assigning each sample to a positive or negative class with a 50\% probability (uniform), assigning classes based on the overall class distribution (class proportion), or assigning all samples as negative (all negative). Performance metrics for random guessing predictions were calculated by averaging the results over 10'000 random simulations.}
    \label{tab:prediction_metrics}
    \end{subtable}
\end{table}

%%======================================================================%%
%% Methods %%
\clearpage
%TC:ignore
\section{Methods}\label{methods}
\subsection{Retrieval of metadata and amplicon sequences}\label{data_fetch}

We retrieved 16S rRNA gene amplicon sequences and associated metadata from the DIABIMMUNE study \cite{vatanen2019} from the NCBI Sequence Read Archive (SRA) \cite{kodama2012, leinonen2011} using q2-fondue \cite{ziemski2022} with the Bioproject ID PRJNA497734. Additional metadata particular to the individual 3 subcohorts of the DIABIMMUNE study were fetched from the supplementary materials of the respective publications \cite{yassour2016,vatanen2016,kostic2015}.

\subsection{Metadata processing}\label{meta_proc}

The metadata was parsed in Python (v3.9.19) using numpy (v1.26.4, \cite{harris2020}) and pandas (v2.2.2, \cite{mckinney2010, team2024}), ensuring a consistency of metadata features across subcohorts. Each microbial sample was linked to host characteristics and information on the last postnatal antibiotics exposure (for a detailed listing of all metadata features, see the Data Dictionary in \Cref{supptab:datadict}). Each antibiotics exposure event was described by the duration of the antibiotics taken, the type of antibiotics, and the symptoms causing the antibiotics treatment. The antibiotics were grouped by type and symptoms, causing the antibiotics treatment as depicted in \Cref{supptab:abxinfo}.

\subsection{Amplicon sequence processing}\label{seq_proc}

The raw amplicon sequences were processed in Python (v3.9.19) using the QIIME~2 microbiome bioinformatics platform (v2024.5, \cite{bolyen2019}). For each subcohort adapter trimming was performed using q2-cutadapt \cite{martin2011} and denoising was conducted with q2-dada2 \cite{callahan2016} using truncation lengths specific to the subcohort. The resulting amplicon sequence variants (ASVs) were closed-reference clustered against the SILVA v138.1 V4 reference database \cite{quast2013,pruesse2007}, obtained with rescript \cite{robeson2021}, at a 97\% sequence identity threshold using q2-vsearch \cite{rognes2016}. From the resulting operational taxonomic units (OTUs), samples with fewer than 1000 sequences were removed. The OTUs were taxonomically classified using a Naive-Bayes classifier trained on the SILVA v138.1 V4 reference database using q2-feature-table and q2-feature-classifier \cite{bokulich2018}. The taxonomic classification was used to remove mitochondrial sequences.

The alpha diversity Faith PD \cite{faith1992} per sample was calculated by repeating rarefaction at a sequence depth of 1000 sequences 500 times with q2-diversity \cite{halko2011}, calculating the metric and averaging the metrics across repetitions \cite{schloss2023}. Thereby, a phylogenetic tree was used, which was inferred from the SILVA v138.1 full-length reference database using FastTree with q2-phylogeny \cite{price2010}.

\subsection{Temporal resolution of sequences and antibiotic metadata}\label{sec:DataResolutionProblem}

The gut microbiome samples (and thus alpha diversity measurements) are recorded at daily resolution, whereas antibiotic exposure metadata are available only at half-month resolution, with undocumented rounding conventions. This temporal discrepancy creates analytic uncertainty, as we cannot determine whether a microbiome sample collected within the same half-month bin as an antibiotic administration was obtained before or after the initiation of treatment.
In both cases the observation can either be in the $0$-bin or the $-1$-bin of the plots in \Cref{fig:score,fig:scoresplit}. Consequently, the $0$-bin may contain samples that were collected prior to antibiotic administration, making it impossible for these particular samples to contain any antibiotic-induced effects. In \Cref{fig:score}a, alpha diversity measurements are floored to half-month resolution, causing observations within the same half-month bin as an antibiotic exposure to appear synchronized with the treatment. 
Our analysis would improve if we had access to the precise starting (and ending) day of the antibiotic treatments.

\subsection{Details for synthetic dataset}\label{sec:Details for synthetic dataset}
For more details on the synthetic dataset, model training, and insights from it, beyond the information provided in this section, see \citet{chardonnet2023anomaly}.

\subsubsection{The synthetic base data model}\label{sec:The synthetic base data model}
The synthetic base dataset is defined as the solution of the stochastic differential equation (SDE) 
\begin{align}
\begin{split}
    dX_{t} &= -\theta(X_{t}-m(t))dt + \sigma dW_{t}, \\
    X_{0} &= x_{0},
\end{split} \label{eq:generating_process}
\end{align}
which admits a unique strong solution \citep[Thm. 7, Chap. V]{Pro1992} and follows the drift function $m$ with Brownian noise.
Here, $m : [0,T] \to \mathbb{R}$ is a bounded continuous function, $W$ is a $1$-dimensional Brownian motion, and $\theta,\sigma\in\mathbb{R}$ are positive. The solution of \eqref{eq:generating_process} is a unique stochastic process $X$ from which we can sample multiple different paths. For each realization of the Brownian motion $W$, we obtain a realization of $X$.

\paragraph{Conditional distribution}
The conditional distribution of $X_{t}|\mathcal{A}_{t}$, where $\mathcal{A}_t = \mathcal{A}_{\tau(t)}$ is the $\sigma$-algebra generated from all observations made until the last observation time $\tau(t)$ before the current time $t$, is Gaussian, with conditional expectation and variance given by
\begin{align}
    \mathbb{E}[X_{t}|\mathcal{A}_{\tau(t)}] & =e^{-\theta (t-\tau(t))}X_{\tau(t)} + \int_{\tau(t)}^{t}e^{-\theta (t-s)}\theta m(s)ds ,\label{eq:true cond exp} \\ 
    \text{Cov}(X_{t}|\mathcal{A}_{\tau(t)}) &= \int_{\tau(t)}^{t}e^{-\theta (t-s)}\sigma\sigma^{T} e^{-\theta^{T} (t-s)}ds. \label{eq:true cond var}
\end{align} 
This formulation also applies in a multi-dimensional setting.
We note that conditional expectation and variance together determine the entire law of the Gaussian conditional distribution.

\paragraph{Specifics of the base dataset}
We define $m$ as a neural network (with two layers of size 16 and ReLU activation) with random weights taking only $\cos(2\pi t / \mathcal{T})$ and $\sin(2\pi t / \mathcal{T})$ as input and note that we use the same function $m$ for all generated datasets. 
Hence, $m$ is $\mathcal T$-periodic, therefore bounded.
We use $\mathcal{T}= T/2$ and $T=1$.
Additionally, we set $x_0 = m(0)$ and the parameters of the SDE $\theta=15$ and $\sigma=0.3$.
Solutions of the SDE \eqref{eq:generating_process} are sampled using the Euler-Maruyama scheme on a fixed equidistant grid with $401$ grid points (i.e., step-size $\delta = 1/400 = 0.0025$). On the training set, each grid point was used as an observation with probability $0.1$ (such that the model learns to predict also further into the future), while all grid points were used as observations in the test sets.
The data was rescaled linearly to approximately fit the desired range (between 0 and 1).

\paragraph{Training Details}
We used $N=80K$ paths for the training set and $20K$ paths as validation set.
The NJODE model was trained for $50$ epochs with learning rate $0.001$ and batch size $200$. Regarding the NN architectures of the NJODE, we had $2$ hidden layers with bias for each of the three networks (jump, neural ODE, and readout), with the configuration as in \Cref{nn_config}.
During training, we use a dropout rate of $0.1$ for each NN and layer, and we use (a posteriori) early stopping based on the validation loss to retrieve the best model state.
The model only got the observations of the synthetic model as input $X$, without any covariate process $C$.
\begin{table}[!h]
    \begin{tabular}{| c | c | c |}
    \hline
     & layer 1 (neurons, activation) & layer 2 (neurons, activation) \\
    \hline
     Jump NN & 200, tanh & 200, tanh \\ 
     \hline
     neural ODE & 300, tanh & 300, ReLU   \\
     \hline
     Readout NN & 200, tanh & 200, tanh   \\
     \hline
    \end{tabular}
\caption{\label{nn_config} NN configuration for the synthetic datasets.}
\end{table}

\subsubsection{Injected anomalies}\label{sec:Injected anomalies}
We inject the following types of anomalies into the base model to generate the anomalous datasets:
\begin{itemize}
    \item change of \textbf{drift}: for a random time interval $[u,v] \subset [0,T]$, the drift function is set to some value $\tilde m$, i.e., $m(t) := \tilde m$ for $t \in [u,v]$;
    \item change of \textbf{diffusion}: for a random time interval $[u,v] \subset [0,T]$, the diffusion coefficient $\sigma$ is replaced by $\tilde \sigma$;
    \item additional \textbf{noise}: for a random time interval $[u,v] \subset [0,T]$, white noise is added to the process, i.e., $X_{t}$ is replaced by $X_{t} + \epsilon_{t}$ with independent $\epsilon_{t}\sim\mathcal{N}(0,\sigma_\epsilon^2)$ for every grid point $t$ in $[u,v]$;
    \item \textbf{spikes}: at random time points $\mathscr C= \{c_1, \dotsc, c_k\}$ a random spiking value $v_t$ is added with random sign $s_t \in \{+1,-1\}$, i.e., $X_t$ is replaced by $X_{t} + s_t v_t$ for $t \in \mathscr C$.
\end{itemize}
Each timestamp gets an anomaly label $y_t$ in $\{0,1\}$, where $1$ corresponds to an anomaly. In particular, we set $y_t=1$ if $t \in [u,v]$ or $t \in \mathscr C$, respectively, and $y_t=0$ otherwise.

\begin{figure}[hp]
     \centering
     \includegraphics[width=\linewidth]{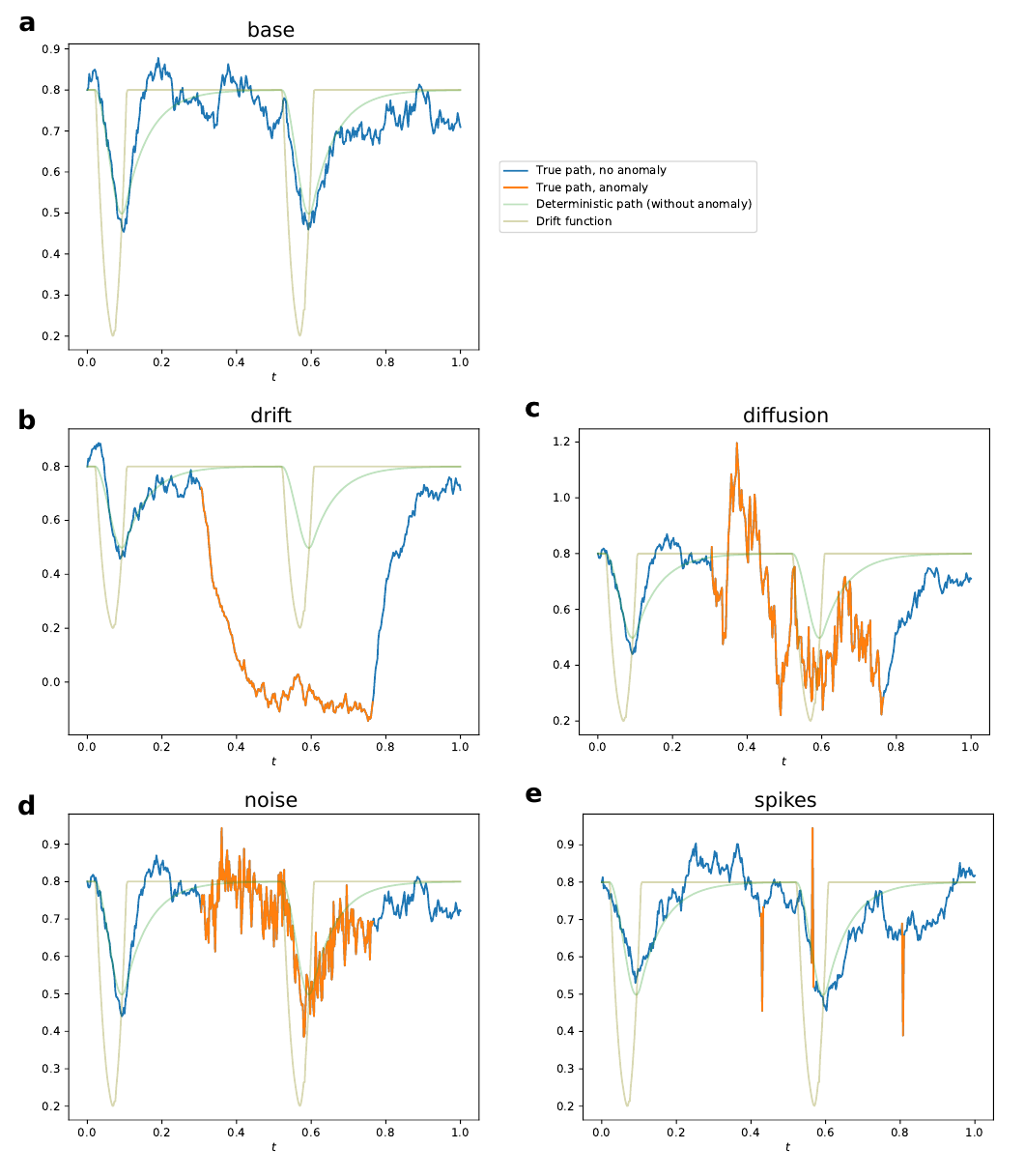}
        \caption{Samples of the synthetic Ornstein-Uhlenbeck based process: anomaly-free base process (\textbf{a}) and different anomalous versions (\textbf{b-e}). 
        The final path is in blue (no anomaly) and orange (anomaly). The deterministic path (of the SDE without diffusion term) without anomaly is in green and its corresponding (anomaly-free) drift function $m$ in olive green.}
        \label{suppfig:anomaly types}
\end{figure}

\paragraph{Specifics of the anomalous datasets}
For drift, diffusion, and noise anomalies, we select a random range $r\sim\mathcal{U}([0.1,0.6])$, which is the proportion of the time series that is anomalous. The starting time of the anomaly is set uniformly at random over the entire time interval. 
For the drift anomaly, we set $\tilde m = 0$, which lets the anomalous samples approach $0$ within the anomaly region.
For the diffusion anomaly, we set the coefficient $\tilde \sigma = 5 \sigma$ and for the added noise we use $\sigma_\epsilon = 0.05$.
For the spike anomaly, each grid point is added to $\mathscr C$ independently with probability $0.005$, $s_t \sim U(\{-1,1\})$ and $v_t \sim\mathcal{U}([0.2,0.5])$.
Generated samples of the anomaly-free and anomalous datasets are shown in \Cref{suppfig:anomaly types}.

\paragraph{Training of anomaly detection framework}
For the synthetic anomalous datasets with dense observations, we use the aggregated scores
\begin{equation*}
    S^{\text{ag}}_t = \sum\limits_{l=-L}^{L}\sum\limits_{k \in \mathcal{K}}w_{lk} S_{ t + \delta l, t + \delta (l-k)},
\end{equation*}
where $L\in\N$ controls how many neighboring scores are considered and $\mathcal K \subset \N$ corresponds to the different $k$ step-ahead predictions. 
In general, this formulation allows to use future scores, hence, future information of the path, to decide whether there is an anomaly at $t$. Depending on the context, if this is not feasible, then only past scores ($l \leq 0$) or only current scores ($l = 0$) could be used.
Whether the usage of future and past scores is necessary depends on the type of anomaly. For example, $l=0$ should be enough for the spike anomaly, since the anomaly can be detected by comparing the current value to the expected current value. On the other hand, for example, the noise anomaly can have a small added noise at $t$, which is not reliably detectable as an anomaly alone, so neighboring scores are crucial to decide whether one is in the anomalous region.
Hence, we train a different anomaly detection (AD) module, i.e., different aggregation weights $w_{lk}$ for the aggregated score $S_t^{\text{ag}}$ for each type of anomaly. 
We further restrict the weights through the factorization 
$$ w_{lk} := \sigma(b_l \,a_k),$$
where $\sigma$ is the sigmoid or logistic function and $a_k\in \R$, $k \in \mathcal{K}$ and $b_l \in \R$, $-L \leq l \leq L$ are the raw trainable weights.
The module's predicted probability of an anomaly is $\sigma(S_t^{\text{ag}})$. We fix $L=5$ and $\mathcal K= \{1,3,5,7,10,12,15,18,20\}$ and use stochastic gradient descent (SGD) with Adam to train the aggregation weights with the cross-entropy loss to correctly classify the anomaly labels $y_t$ (1 if anomalous, 0 otherwise) at each feasible sampling grid point. The feasible grid points are those, where all necessary scores are available. With our choice of $L$ and $\mathcal{K}$, this means all 401 grid points except for the first 25 and the last 5 ones, leaving 371 feasible grid points for each sample.
For each type of anomaly, we use an aggregation training set with $N=800$ and a test set with $200$ samples. Each AD module is trained for $50$ epochs with learning rate $0.01$, batch size $100$, and additional $L^2$-weight regularization with factor $\lambda=1$. We did not use early stopping.
For anomaly classification, we used the threshold $0.5$ without further optimization, yielding good results for all anomaly types. %However, we note that optimization of the threshold could lead to further improvements, in particular, if the objective is to reduce either the type I (false positives (FP)) or type II (false negatives (FN)) error.
However, further improvements could be realized by optimizing the threshold based on the relative costs of Type I (false positives (FP)) and Type II (false negatives (FN)) errors.
The test set is only used for plotting the results of \Cref{fig:simulation_results}b, where we see good detection of all types of anomalies. Moreover, inspecting the false positives and false negatives, one can mostly infer why they were misclassified. For example, the two false positive spike anomalies occur at times where the anomaly-free path has large increments.

\paragraph{Quantitative evaluation of anomaly detection framework}
In addition to the qualitative evaluation of the anomaly detection framework in \Cref{fig:simulation_results}b, we report aggregated quantitative results in \Cref{supptab:simulation_results_quantitative}.
We use additional independent evaluation sets with $1500$ samples for each type of anomaly, where we test the classification quality of the AD modules by extracting standard statistics. Furthermore, we test the AD modules on an independent anomaly-free dataset (i.e., all grid points of all samples have label $0$) of $1500$ samples, where we report the false positive rate (i.e., 1 minus recall of label $0$) only. Each evaluation dataset has a total of $1500*371=556.5K$ labels to be predicted, and the supports of label $1$ are $210'615$ (drift anomaly), $206'276$ (diffusion and noise anomaly), and $2'767$ (spike anomaly).
For all anomaly types, we have a recall of label $1$ of at least $97\%$, meaning that the AD module detects nearly all anomalous labels, while having a false positive rate of less than $9\%$ on the anomaly-free dataset. In particular, for the spike anomaly (which is the most relevant for our real-world dataset), the AD module detects all anomalous labels and only misclassifies $2\%$ of the labels of the anomaly-free dataset; the precision and recall of label 1 are small due to the large class imbalance with approximately $3K$ positive compared to $554K$ negative labels.

\subsubsection{Ablation study: decreasing the size of the training set}\label{sec:Ablation study: decreasing the size of the training set}
Since our real-world dataset has very limited size, we tested the influence of the size of the training set on the predictive performance of the NJODE model within our synthetic dataset. 
In particular, we retrained the same model (that was trained with $80K$ paths for the synthetic anomaly detection result) with $N =200$ paths. 
In the training, the number of epochs was increased in order to have the same amount of forward passes, i.e., $20K$ epochs for the $200$ paths.
Otherwise, the specifics of the dataset and model training were the same as before.

To quantify the predictive quality of the resulting trained NJODE models, we compared their predictions with the ground truth of the conditional moments (\Cref{eq:true cond exp,eq:true cond var}).
Concretely, we computed the mean square error (MSE) between the predictions and the truth at each observed timestamp before and after jumps. We simply used the Euler method whenever the computation of the ground truth involved an integral. For the second moment, the MSE was calculated using the conditional standard deviation by applying the square root to the conditional variance. In case the prediction of the conditional variance was non-positive, it was simply replaced by $10^{-4}$. The model trained with $80K$ paths was evaluated during training after each epoch, while the evaluation was performed every $400$ epochs for the model trained on $200$ paths. In both cases, we performed the evaluation on the same $N_{test}=200$ paths. We then reported the best evaluation (in MSE) over the whole training, together with the corresponding standard deviation (of the square error) with the notation: $\operatorname{MSE}\pm \operatorname{STD}_{\operatorname{MSE}} / \sqrt{N_{test}}$. The evaluation is reported for both the conditional expectation and variance separately.

The results are reported in \Cref{result_diff_dataset_sizes}. Reducing the size of the training set by the factor $400$ yields an increase by a factor of approximately $2.4$ and $2.4^2 = 5.76$ in the MSE of the conditional expectation and standard deviation, respectively. As expected, we see a clear decline in performance with the smaller training set. Nevertheless, the results are still on comparable scales, which is promising for our application on scarce real-world data. 
Although the theoretical results of \citet{krach2022optimal} imply convergence as the number of training paths increases, the convergence rate strongly depends on the complexity of the dataset. Hence, we only have a very limited understanding of the quantitative implications of small data for real-world datasets (since we would need more data to reasonably analyze it). Ultimately, we can only empirically test whether the trained model can be used successfully, in our case for anomaly detection.
\begin{table}[!h]
    \begin{tabular}{| c | c | c |}
    \hline
    dataset size~$N$ & cond. exp. MSE ($\times 10^{-4}$) & cond. std MSE ($\times 10^{-4}$) \\
    \hline
     80K & $1.058 \pm 0.2679$ & $ {0.4902\pm 0.09143}$ \\ 
     \hline
     200 & $2.429\pm 0.5053$ & $2.7484 \pm 0.3038$   \\
     \hline
    \end{tabular}
    \caption{\label{result_diff_dataset_sizes}Results comparing the NJODE model trained with different dataset sizes.}
\end{table}
% TO CHANGE

\subsection{Details for the NJODE model}\label{sec:Details for the NJODE model}

\subsubsection{Microbiome data description}\label{sec:Microbiome data description}
The target process $X$ (in the case of our real-world dataset, the alpha diversity metric) and the process of covariates $C$ (in this case, the delivery mode and the changing milk diet and weaning) are assumed to be continuous-time stochastic processes on a finite interval $[0,T]$. Even though the delivery mode is a static random variable, we can view it as a stochastic process with constant paths. Each of the $N$ patients corresponds to an identically distributed and independent (iid.) copy $Z^{(i)} = (X^{(i)},C^{(i)})$ of these target processes. While these processes are defined in continuous-time, we only have discrete observations of them at a random number $n^{(i)}$ of random observation times $t_1^{(i)} < \dotsb < t_{n^{(i)}}^{(i)}$. With $\tau(t) = \max\{t_k | t_k \leq t \}$ we denote the last observation time before (or at) time $t$.
The timestamps for the data are counted in days, with a maximum sample length of $T=1162$ days. When training the NJODE model, we transform the time to correspond to the interval $[0,1]$, since this improves the model quality. In particular, we divide any real timestamp given in days by $1162$ to get the model time. Depending on the part of the anomaly framework, we will either use the real timestamp or the model timestamp, which should be clear from the context.
Even though observations are always complete in this dataset, i.e., all coordinates of $Z$ are observed, we use a masking process $M$ to mask some observations of $X$ during training to learn the long-term multistep ahead predictions (see the next paragraph for more details).
We note at this point that 1-step ahead refers to predicting from each observation until the next observation (not only for one model time step), while multistep ahead means to predict multiple observations ahead.

Training the NJODE model on the train set (all samples without antibiotic events) effectively conditions it to predict the distribution of $X$ without exposure to antibiotics; hence, such exposure should be detected as anomaly. 
The anomaly we expect to see after antibiotics administration is a downward jump of the alpha diversity. In contrast to a spike (as in one of our synthetic anomalous datasets), the path continuous from the value after the jump and does not jump back up. However, from the perspective of the anomaly score $S_{t_i, t_j}$ for detecting this anomalous behavior, this is similar to a spike at $t_i$, if $t_j$ was (shortly) before and $t_i$ (shortly) after antibiotic administration. In particular, just seeing the two values at $t_j$ and $t_i$, a downward jump and a spike are indistinguishable.  
If more observations are available, a jump should, in general, be easier to detect than a spike, since multiple observations in a row are different than expected.

\subsubsection{The Signature transform}\label{sec:signature}
    Let $J$ denote a closed interval in $\R$.
    Let ${X}:J\rightarrow\R^{d}$ be a continuous path with finite variation.
    The signature of ${X}$ is defined as
    \begin{equation*}
        \operatorname{Sig}({X}) = \left(1, {X}_J^1, {X}_J^2, \dots\right),
    \end{equation*}
    where, for each $m\geq 1$,
    \begin{equation*}
        {X}_J^m = \int_{\substack{u_1<\dots<u_m \\ u_1,\dots,u_m\in J}} d{X}_{u_1}\otimes\dots\otimes d{X}_{u_m} \in (\R^d)^{\otimes m}
    \end{equation*}
    is a collection of iterated integrals.
    The map from a path to its signature is called signature transform.
In practice, we are not able to use the full (infinite) signature, but instead use a truncated version.
    The  truncated signature of ${X}$ of order $m$ is defined as 
    \begin{equation*}
        \pi_m({X}) = \left(1,{X}_J^1,{X}_J^2,\dots,{X}_J^m\right),
    \end{equation*}
    i.e., the first $m+1$ terms (levels) of the signature of ${X}$.
Note that the size of the truncated signature depends on the dimension of ${X}$, as well as the chosen truncation level.
Specifically, for a path of dimension $d$, the dimension of the truncated signature of order $m$ is given by
\begin{equation}\label{eq:sig_nb_terms}
\begin{cases}
m+1, & \text{if } d =1, \\
 \frac{d^{m+1}-1}{d-1}, & \text{if } d >1. 
\end{cases}   
\end{equation}
When using the truncated signature as input to a model this results in a trade-off between accurately describing the path and model complexity.
A good introduction to the signature transform with its properties and examples can be found in \citet{Chevyrev2016APO, KiralyOberhauser2019, fermanian2020embedding}, and more precise explanations of its usage in the context of NJODEs is provided in \citet{krach2022optimal}. 

\subsubsection{Model setup \& Training} 
We use a NJODE model \eqref{equ:PD-NJ-ODE} with different input and output variables \citep{heiss2024nonparametricfilteringestimationclassification}. 
As inputs to the model we use $X$ and additionally the covariates process $C$, however, the signature is only computed with $X$.
The output $Y = (Y^1,Y^2,Y^3)$ consists of predictions $(Y^1,Y^2)$ of the first two moments of the alpha diversity process $X$, and a direct prediction $Y^3$ of the conditional variance. 
We use the input-output loss function \citep{heiss2024nonparametricfilteringestimationclassification} for the moment predictions $(Y^1, Y^2)$, scaled with weights $\gamma_i$ for the loss of $Y^i$, $i=1,2$.
The conditional variance output is trained to minimize the squared distance to the two terms
\begin{equation*}
    Y^2_t - (Y^1_t)^2 \quad \text{and} \quad (X_t - (Y^1_t))^2.
\end{equation*}
If the moment predictions $(Y^1,Y^2)$ are optimal, i.e., replicating the respective conditional variances, then the first term is exactly the conditional variance, i.e., 
\begin{equation*}
    Y^2_t - (Y^1_t)^2 =  \E[X_t^2 \, | \, \mathcal{A}_{\tau(t)} ] - \E[X_t \, | \, \mathcal{A}_{\tau(t)} ]^2 = \operatorname{Var}[X_t \, | \, \mathcal{A}_{\tau(t)}] .
\end{equation*}
Due to numerical errors, this is not a reliable estimator. In particular, it does not always satisfy the non-negativity constraint, hence we do not use it directly, but instead train $Y^3$ with it. Training $Y^3$ only with this target should lead to perfect reconstruction (since $Y^1,Y^2$ are model outputs), which does not resolve the problem. Hence, we additionally train $Y^3$ with the second term as target. Here, if $Y^1$ is optimal, then 
\begin{equation*}
     (X_t - (Y^1_t))^2 = (X_t - \E[X_t \, | \, \mathcal{A}_{\tau(t)} ] )^2.
\end{equation*}
Similarly as in the proof of the standard NJODE \citep{krach2022optimal}, the minimizer of the objective function
\begin{equation*}
    Z \mapsto \E\left[ \frac{1}{n} \sum_{i=1}^n  \left\lvert (X_{t_i} - \E[X_{t_i} \, | \, \mathcal{A}_{t_{i-1}} ])^2 - Z_{t_{i}-}  \right\rvert_2 ^2 \right]
\end{equation*}
is given by $Z_t = \E[(X_t - \E[X_t \, | \, \mathcal{A}_{\tau(t)} ])^2  \, | \, \mathcal{A}_{\tau(t)} ] = \operatorname{Var}[X_t \, | \, \mathcal{A}_{\tau(t)}]$. Therefore, training $Y^3$ using the second term is consistent to learn the conditional variance.
Moreover, training with the second term has 2 advantages over training with the first one: i) We only use the approximation $Y^1$ but not $Y^2$, hence the numerical error in the target should be smaller. ii) While numerical errors can lead to negative values of the first term, the second term, as a squared expression, is always non-negative, hence it satisfies the constraint of the conditional variance.
We can additionally enforce the non-negativity constraint with another loss term. Moreover, we can use a loss term to enforce that $Y^3$ is $0$ after observations.
This leads to the combined loss function
\begin{equation*}
\begin{split}
    L(Y^3_t, Y^3_{t-}) &= \gamma_3 \left|Y^3_{t-} -  Y^2_{t-} - (Y^1_{t-})^2 \right|^2 + \gamma_4 \left|Y^3_{t-} -  (X_t - (Y^1_{t-}))^2 \right|^2 \\
    & \quad + \gamma_5 \min(Y^3_{t-}, 0)^2 + \gamma_6 (Y^3_t)^2
\end{split}
\end{equation*}
for $Y^3$, which is evaluated at observation times $t_k$. 
In the case of incomplete observations, one additionally has to multiply by the observation mask $M_t$. Moreover, the objective function is built as in \citet{krach2022optimal} from $L$, by averaging over all observation times and taking expectations.
Note that $Y_{t-}$ corresponds to the model output before getting new information from the next observation. When computing this loss during training, the terms $Y^1, Y^2$ are detached, i.e., this loss is only used to optimize $Y^3$, where the current values of $Y^1, Y^2$ are provided as constants.  
The scales are set to $\gamma = (1,0.1,0.05,0.1,1,0)$, however, for $(\gamma_3, \gamma_4, \gamma_5)$ we use a linear interpolation throughout the training (depending on the current epoch) between $0$ and the indicated values (i.e., in the first epoch these weights are set to 0, while they are set to $\gamma_i$ in the last epoch). We set $\gamma_6=0$ due to the noise, since for noisy observations, $Y^3$ does not need to jump to $0$ at observations. For applications with no or much smaller measurement noise, $\gamma_6>0$ can increase data efficiency.

We are not only interested in predicting the conditional expectations at any time $t$ given all information up to $t$, but also in the predictions at time $t$ given only partial information about the covariates $C$ after time $s \leq t$. This corresponds to long-term predictions with incomplete observations.
Therefore, we train our model with the training framework of \citet{krach2024} for long-term predictions, where the probability $p$ of using the full observation as input decreases with the training epoch $e$ as $p(e) = 1 - \max((\min(e, 5000) - 2000)/6000, 0)$, where the model is trained in total for $6000$ epochs.

For early stopping of the model training, we use a validation loss, with scales $\gamma_{\text{val}} = (1,1,0,1,0,1)$ that incorporates long-term prediction in the following way. We use stopping times $s \in \{0.2, 0.4, 0.6, 0.8\}$ in the model time interval $[0,1]$, at which we stop using $X$ as input. Afterwards, we continue the prediction for $\Delta t \approx 0.315$ (corresponding to $366$ days) and we compute the loss using all predictions of the model in the time interval $[0, s+\Delta t]$. Finally, we take the average over the losses computed for the different values of $s$. The early-stopping-based model selection first trains all models for $6000$ epochs and then uses the model at the epoch with minimal validation loss.

Prior to the standard model training, we pre-train the encoder $\rho_{\theta_2}$ and decoder $g_{\theta_3}$ of the model on a randomly generated training set to initially ensure that $g_{\theta_3} \circ \rho_{\theta_2} \approx \text{id}$. 

\subsubsection{Architecture \& Hyperparameters.}
An initial manual hyperparameter optimization led to the following architecture.
The latent dimension (i.e. the dimension of $H$) is set to $d_H = 300$, for the encoder  $\rho_{\theta_2}$ we use a 2 hidden layer neural network with $200$ and $300$ nodes, for the decoder $g_{\theta_3}$ we also use 2 hidden layers with the inverse sizes of $300$ and $200$ nodes and for the neural ODE $f_{\theta_1}$ we use 2 hidden layers with $300$ nodes each; all layers use $\tanh$ activation functions. 
The time of observations is transformed to be in the interval $[0,1]$. In particular, the last observation is at $1162$ days, which we transform to correspond to $T=1$ (i.e., model time is real time in days divided by $1162$).
For solving the neural ODE we use the basic Euler scheme (as in \citet{krach2022optimal}); however, we use larger time-steps corresponding to 7 days instead of the standard time step $\Delta t$ that corresponds to 1 day. Whenever observations are in between, a smaller time step until the next observation is used. This makes the training and inference faster and leads to faster convergence and better results.
We train the model with the Adam optimizer, using a learning rate of $0.001$, a mini-batch size of $30$, and dropout with rate $0.1$. Further implementation-specific details can be found in \citet[Appendix~D]{krach2022optimal}.

We train 4 versions of the model with or without using the signature $\pi_m (\tildeXle{\tau(t)} )$ and with or without using an RNN-type architecture, i.e., using (or not) the latent variable $H_{t-}$ as input to the encoder $\rho_{\theta_2}$. In \eqref{equ:PD-NJ-ODE}, the full model using both options is displayed; for the other versions, the respective components are removed as inputs. Both the signature and the RNN architecture are methods to allow for path-dependence and it is problem/dataset specific, which combination works best \citep[see][Sec.~8.3]{krach2022optimal}. In our case, using the RNN architecture without signatures led to the best validation loss; therefore, we chose this architecture for further evaluations.

\subsection{Details for anomaly detection}\label{sec:Details for anomaly detection} 
At any time $t$ and for any $s \le t$, the NJODE can predict the conditional moments of $X_t$ given all measurements that were available until time $s$. Varying $s$ for fixed $t$, we can change how far the model predicts ahead (i.e., $t-s$). This is useful since an observation at $t$ can seem anomalous from the perspective of $s_1 <t$, while it does not from the perspective of $s_2 \in (s_1,t)$ (e.g., if the anomalous event happened within $(s_1,s_2)$ and the process stabilizes on a different level afterwards). We will use different values for $s$ to analyze the measurements given information about different amounts of antibiotic exposures. 
Moreover, we can decide to use only the covariates $C$ as inputs after $s$, since the model can deal with missing values.
In the following, we leave away $s$ whenever any $s<t$ is suitable and only specify it when necessary.
The NJODE's estimates of the conditional moments (or rather of $\mu_t,\sigma_t^2$), can be used to match a distribution of a prespecified family by estimating its parameters via the method of moments \citep[see][Sec. 0.2.3]{KrachPhDThesis}. We note that this leads to an estimate of the conditional distribution, since we use the conditional moment estimates. For example, if we choose the family of normal distributions, then the estimated $\mu_{t,s},\sigma_{t,s}^2$ directly specify the estimated distribution at $t$ conditioned on the past observations until $s$.
The conditional variance corresponds to the aleatoric uncertainty (of the process $X$), which usually grows with $t-s$. However, the epistemic uncertainty also grows with the forecasting horizon $t-s$ since (i) due to the larger aleatoric uncertainty, the estimation of the conditional mean becomes noisier; (ii) we usually have fewer training observations for large forecasting horizons; (iii) the estimation for long forecasting horizons is harder to learn and there is less focus on it during training. Moreover, these errors cumulate over time, reinforcing the effect\footnote{Future work could incorporate more principled estimation of epistemic uncertainty \citep{azizi2025clearcalibratedlearningepistemic}. One could also consider flagging observations of high epistemic uncertainty to emphasize that such observations cannot be reliably classified as not-anomalous.}. Hence, for large $t-s$, the predicted conditional standard deviation $\sigma_{t,s}$ will underestimate the variance in the observations, since it does not account for the epistemic uncertainty. To account for this, we fit scaling factors (SF) $\alpha_{\text{sf}}(t-s)$, s.t.\ the resulting empirical standardized conditional distributions of observations, when using $\sigma_{t,s} \cdot \alpha_{\text{sf}}(t-s)$ as standard deviation, match the theoretical standardized distribution well (see \Cref{sec:Details for the computation of scaling factors} for more details). 
Continuing the example of a normal distribution, we fit the SFs s.t.\ the empirical distribution of the rescaled z-scores $\tilde z^{(i)}_{t,s} = \frac{x^{(i)}_t - \mu_{t,s}^{(i)}}{\sigma_{t,s}^{(i)} \cdot \alpha_{\text{sf}}(t-s)}$ for several observations $x^{(i)}$ is close to standard normal.
    
Let $F_{t,s}$ be the cumulative distribution function of the estimated conditional distribution (using the fitted scaling factors) at $t$ given observations until $s$, then the left-sided p-value of a new observation $x^{(i)}_t$ can be computed as $p^{(i)}_{t,s} = F_{t,s}(x^{(i)}_t)$ and similarly for other p-values. To allow for a better distinction of small p-values (those representing anomalies), we transform them to scores as $S_{t,s}^{(i)} = - \ln(p^{(i)}_{t,s})$, where larger values correspond to higher anomalies.
These scores can be used to describe abnormalities (whether our model detects it and how large it is), or to predict and classify whether an observation is abnormal.

\subsection{Details for the computation of scaling factors}\label{sec:Details for the computation of scaling factors}
\begin{figure}[hp]
    \centering
    \includegraphics[width=\linewidth]{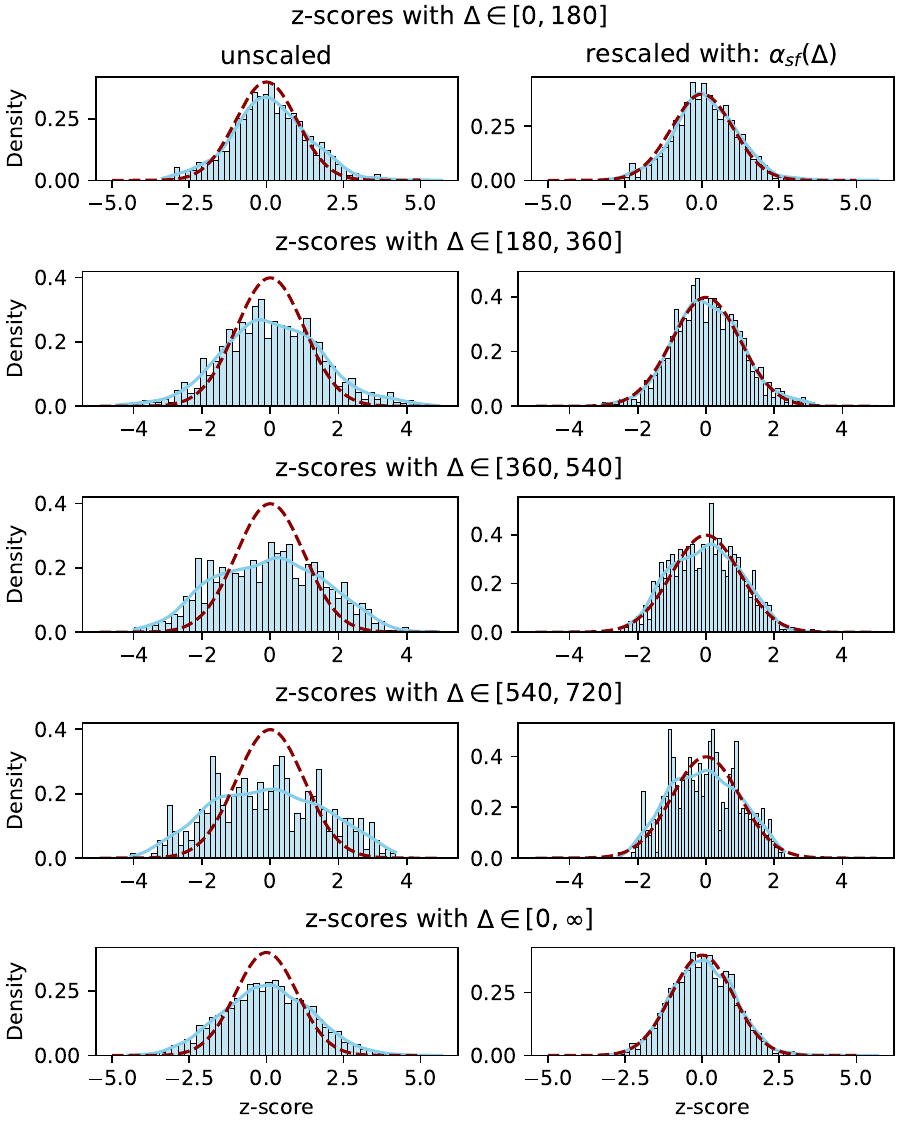}
    \caption{Histograms of unscaled (left) and rescaled (with $\alpha_{\text{sf}}(\Delta)$) z-scores (right) for different intervals of days since last observation $\Delta=t_{k_2}-t_{k_1}$ for $(k_2-k_1)$-step-ahead predictions for all observed combinations $t_{k_1}<t_{k_2}$. We see a good fit of the resulting rescaled z-scores to the standard normal distribution.}
    \label{suppfig:histogram zscores with sf overview}
\end{figure}

To compute scaling factors (cf.~\Cref{sec:The anomaly framework design,sec:Details for anomaly detection}) for long-term / multi-step ahead variance predictions, we first compute z-scores $z_{t_k,t_j}^{(i)} = \frac{x^{(i)}_{t_k} - \mu_{t_k,t_j}^{(i)}}{\sigma_{t_k,t_j}^{(i)}}$ with all different combinations of sorted pairs of observation times $t_j < t_k$, corresponding to $\Delta = t_k - t_j$ days since their last observation. 
We would like to find the scaling factors $\alpha_{\text{sf}}(\Delta)$ for any $\Delta \in \N$ such that the scaled z-scores $\tilde z_{t_k,t_j}^{(i)} = \frac{x^{(i)}_{t_k} - \mu_{t_k,t_j}^{(i)}}{\sigma_{t_k,t_j}^{(i)} \cdot \alpha_{\text{sf}}(t_k-t_j)}$ have a standard normal distribution. The choice to use multiplicative scaling factors, instead of, e.g., additive ones, is a modeling decision. The advantage of multiplicative scaling factors is that they are easy to compute, since one can simply use the empirical standard deviation of the samples z-scores.
However, there is no closed-form solution to do this; hence, it boils down to an engineering task to find good scaling factors. 
In our real-world microbiome dataset, we have a total of 4122 z-scores using all different $t_j < t_k$ pairs for all samples in the validation set. Hence, we do not have enough samples for each $\Delta \in \N$ to reliably compute the scaling factor $\alpha_{\text{sf}}(\Delta)$ only with those z-scores having exactly $\Delta$ days since their last observation. Therefore, we compute them in moving windows. For $\Delta < 60$, we set $\alpha_{\text{sf}}(\Delta) = 1$, since we have too little data for reliable computations and since we empirically see that this leads to a good distribution fit. For any $\Delta \in \N_{\ge 60}$, we consider all z-scores $z_{t_k,t_j}^{(i)}$ for which $t_k - t_j \in [\Delta - 60, \Delta + 60]$ and compute their non-centered standard deviation\footnote{The non-centered standard deviation corresponds to the square root of the 2nd empirical moment $\sqrt{\E[Z^2]} = \sqrt{\text{Var}(Z) + \E[Z]^2}$. We use this instead of the standard deviation, since it also accounts for the epistemic uncertainty in the computation of the mean, since we already subtract the conditional mean $\mu_{t_k,t_j}^{(i)}$ in the definition of $z_{t_k,t_j}^{(i)}$.}. To further smoothen the resulting values, we use the moving average over $10$ days (MA(10)). Then, we define $\alpha_{\text{sf}}(\Delta)$ as the cumulative maximum over these MA(10) values, which implements our belief that the epistemic uncertainty grows with $\Delta$. These computations are shown in \Cref{suppfig:inf_distribution}b. Moreover, the comparison of the unscaled z-scores $z_{t_k,t_j}^{(i)}$ and rescaled z-scores $\tilde z_{t_k,t_j}^{(i)}$ in different $\Delta$-intervals in \Cref{suppfig:histogram zscores with sf overview} shows a good distribution fit after rescaling. To achieve this, we tried different moving window sizes, moving average values, etc.~in a backward engineering approach.

Hence, we can use the standard normal distribution to compute p-values of the rescaled z-scores $\tilde z_{t_k,t_j}$ on the test set. Or in different words, the conditional distribution of new test observations $x_{t_k}$, given information up to $t_j$, is $N(\mu_{t_k,t_j}, (\sigma_{t_k,t_j} \cdot \alpha_{\text{sf}}(t_k-t_j))^2)$, which can be used to compute p-values of $x_{t_k}$.

\subsection{Modelling set-up for score vs. baseline predictions}\label{sec:predictive_setup}
As opposed to previous sections that relied on multi-step ahead score predictions from the antibiotics exposure onwards to infer dynamics from antibiotic exposures, we used one-step ahead score predictions for the predictive setup, carrying no information leakage about antibiotic administration. In the setting of \Cref{sec:Details for anomaly detection}, we only compared single-step and multi-step ahead scores of post-abx-exposure samples with only one-step ahead scores of pre-abx samples and adjusted for the difference via scaling factors. Here, we only compare one-step ahead scores. Therefore, $t-s$ does not carry any information on antibiotic administration. Ground-truth positive antibiotic events were defined as the first observed microbial sample within three months of an infant’s first antibiotic exposure, given at least one prior microbial sample. Ground-truth negatives were samples from the antibiotics cohort with no known history of antibiotics exposure and at least one prior observed microbial sample. This resulted in 70 positive and 247 negative class samples.

%%===============================%%
%% Data availability %%

\section{Data availability}\label{data_avail}

Microbial amplicon sequences and the associated metadata were obtained from three cohorts of the DIABIMMUNE study \cite{vatanen2019} with the BioProject ID PRJNA497734. Additional metadata particular to the individual 3 subcohorts of the DIABIMMUNE study were fetched from the supplementary materials of the respective publications \cite{yassour2016,vatanen2016,kostic2015}. The feature table used by our anomaly framework is available in the data folder in the GitHub repository at \href{https://github.com/adamovanja/anomaly_microbiome_data_processing}{\url{https://github.com/adamovanja/anomaly_microbiome_data_processing}}.

%%===============================%%
%% Code availability %%

\section{Code availability}\label{code_avail}

The pipeline to fetch and process the microbial sequences and the associated metadata is available in the GitHub repository at \href{https://github.com/adamovanja/anomaly_microbiome_data_processing}{\url{https://github.com/adamovanja/anomaly_microbiome_data_processing}}.
The anomaly framework is available in the GitHub repository at \href{https://github.com/MarkusChardonnet/Probabilistic_forecasting_for_Anomaly_Detection}{\url{https://github.com/MarkusChardonnet/Probabilistic_forecasting_for_Anomaly_Detection}}.

%%======================================================================%%
%% References %%
\bibliography{abx_njode}

%% \bibliography{sn-bibliography}% common bib file


%%======================================================================%%
\backmatter
%% Acknowledgements %%

\section*{Acknowledgements}

We thank Paula Momo Cabrera (ETH Zürich), Emma Slack (ETH Zürich) and Fannie Kerff (ETH Zürich) for insightful discussions on microbiome perturbation dynamics, and Christian L. Müller (Helmholtz \& LMU Munich) for feedback on alternative and prospective modelling approaches. We also thank Alina Ofenheimer (LBI Institut for Lung Health) for the clinical classification of antibiotics and their respective symptom profiles.

%%===============================%%
%% Extended data %%
\FloatBarrier
\newpage
\section*{Extended data}
\beginextdata

\begin{table}[hp]
\centering
\caption{Quantitative evaluation results of the trained anomaly detection framework on independent evaluation sets of the simulated time-series for the different anomaly types and on an independent anomaly-free evaluation set. On the anomalous datasets we report the rounded support in thousands (Sup), Precision (Pr), Recall (Re), and F1 score (F1) for both labels and additionally the micro-averaged F1 score over both labels. On the anomaly-free dataset the false positive rate $\text{FPR} = \tfrac{\# \text{FP}}{\# \text{labels}}$ is reported.
    \label{supptab:simulation_results_quantitative}}

% \small
\begin{tabular}{ l | r r  r  r | r r  r  r | r | r }
        \toprule
                  & \multicolumn{9}{c|}{Anomalous} & \multicolumn{1}{c}{Anomaly-free}                                                                                  \\
        \midrule
                  & \multicolumn{4}{c|}{Label 0}   & \multicolumn{4}{c|}{Label 1}     & \multicolumn{1}{c|}{Micro-av.} &                                               \\
        Anomaly   & Sup                            & Pr                               & Re                             & F1   & Sup & Pr   & Re   & F1   & F1   & FPR  \\
        \midrule
        drift     & 346                            & 0.99                             & 0.93                           & 0.96 & 211 & 0.90 & 0.99 & 0.94 & 0.95 & 0.04 \\
        diffusion & 350                            & 0.99                             & 0.89                           & 0.94 & 206 & 0.85 & 0.99 & 0.91 & 0.93 & 0.08 \\
        noise     & 350                            & 0.98                             & 0.89                           & 0.93 & 206 & 0.84 & 0.97 & 0.90 & 0.92 & 0.09 \\
        spike     & 554                            & 1.00                             & 0.96                           & 0.98 & 3   & 0.11 & 1.00 & 0.20 & 0.96 & 0.02 \\
        \bottomrule
    \end{tabular}
\end{table}

\begin{table}[hp]
    \centering
    \caption{Characteristics of the first, second, and third antibiotic exposures .}\label{supptab:first_second_abx_char}
    \begin{tabular}{lllll}
        \toprule
        \textbf{Exposure} & \textbf{Age at exposure,} & \textbf{Abx duration,} & \textbf{Top 2} & \textbf{Top 2}     \\
        & \textbf{mean ± stddev}    & \textbf{mean ± stddev} &      \textbf{abx types}                    &    \textbf{abx reasons}                            \\
                          & \textbf{[months]}         & \textbf{[days]}        &                          &                                \\
        \midrule
        \textbf{1st}      & 10.10 ± 5.66              & 7.41 ± 4.12            & Penicillin,              & Otitis media,                  \\
                          &                           &                        & Other                    & Infection of respiratory tract \\
        \textbf{2nd}      & 12.69 ± 5.02              & 7.33 ± 2.86            & Penicillin,              & Otitis media,                  \\
                          &                           &                        & Other                    & Infection of respiratory tract \\
        \textbf{3rd}      & 14.47 ± 4.60              & 9.04 ± 8.49            & Penicillin,              & Otitis media,                  \\
                          &                           &                        & Macrolide                & Infection of respiratory tract \\
        \bottomrule
    \end{tabular}
\end{table}

\newpage

\FloatBarrier
\section*{Supplementary information}
\beginsupplement
\section{Extended Literature Review: Deep Learning Alternatives to NJODE}\label{appendix:LiteratureAlternativesToNJODE}
In this subsection, we discuss further deep learning models for forecasting time series.

\textbf{TabPFN.} The foundation model TabPFN-TS \citep{hoo2025tabularfoundationmodeltabpfn} can make zero-shot forecasts for irregularly observed time series based on TabPFNv2 \citep{hollmann2025accurate} (which is the successor of TabPFN \citep{hollmann2023tabpfntransformersolvessmall}). At the time of its publication, TabPFN-TS achieved the first rank on the GIFT-Eval Time Series Forecasting Leaderboard \cite{aksu2024gifteval}. However, a straightforward application of TabPFN-TS would give us a forecast only based on the history of one patient without having learned the general dynamics of the microbiome from other patients. Therefore TabPFN-TS would not be able to make an informed forecast directly after the first measurement of a patient, for example.\footnote{TabPFN-TS is well suited for forecasts based on a longer history of a time series, because TabPFN-TS only receives information on the underlying dynamics from this history. TabPFN-TS does not use any information from other patients. Increasing the number of training patients would not give any benefit to TabPFN-TS, because it does not share any information across patients. In contrast, we train NJODE on many training patients to learn the underlying dynamics. The more training patients we train on, the better we understand the underlying dynamics. This allows us to make reasonable forecasts for a new patient directly after the first measurements. TabPFN-TS has shown surprisingly good performance in many settings. But our setting is different, as we observe many very short time series rather than a few long time series. For the majority of our patients, we have less than 10 observations in total, and we want to have a good forecasting performance directly after the first observation.}
In contrast, NJODE can give us an asymptotically optimal forecast directly after the first measurement of a patient if it has been trained on sufficiently many other patients before.
%In theory, it would be possible to fine-tune TabPFN-TS on the training (i.e., the histories of measurements from other patients). However, the the current API of the proprietary TabPFNv2 does not allow to do so.
We train NJODE on the histories of hundreds of patients to learn the dynamics of the microbiome before we make forecasts for new patients.
It would be interesting future work to fine-tune TabPFN-TS on our training dataset to compare it against NJODE.

\textbf{TiRex.} Recently, TiRex \citep{auer2025tirexzeroshotforecastinglong}, a foundation model similar to TabPFN-Ts, but based on an xLSTM-architecture \citep{beck2024xlstm}, made it to the top of the GIFT-Eval Time Series Forecasting Leaderboard (with a certain degree of test data leakage) \cite{aksu2024gifteval}. It would be interesting future work to fine-tune TiRex on our training dataset to compare it against NJODE.

\textbf{Transformers.} It would be interesting future work to compare NJODE with transformers \citep{NIPS2017_3f5ee243AttentionIsAllYouNeed}. In principle, transforms can also deal with irregularly observed time series via suitably chosen temporal encoding. However, the inductive bias of the method plays a crucial role in our setting, as the number of training patients is very limited. We believe that the inductive bias of NJODEs is more aligned with our implicit prior for microbiome dynamics than the inductive bias of transformers, since NJODEs prefer a simple ODE describing the forecast rather than the forecast itself being simple \citep{NJODE3,HeissInductiveBias2024}. For example, the simplest population models are governed by linear ODEs, which are, in a certain sense, among the simplest ODEs possible, but the solution of a linear ODE is highly nonlinear in $t$. Many neural network architectures extrapolate in a certain sense as linearly as possible \citep{implReg1,implReg2, HeissPart3Arxiv,HeissInductiveBias2024,savarese2019infinite,ongie2019function,williams2019gradient,parhi2022kinds}. For NJODE, the right-hand side of the ODE is parametrized by a neural network. This explains the strong extrapolation performance of NJODE when the optimal forecast follows a linear ODE \citep{NJODE3,HeissInductiveBias2024}, which was experimentally observed in \citet[Appendix~E]{krach2022optimal}.\footnote{Note that NJODE is absolutely not restricted to linear ODEs, since highly non-linear dynamics can be modeled via the nonlinear neural networks. \citep{NJODE3,HeissInductiveBias2024} simply suggest that NJODE's generalization performance is particularly strong if the underlying ODE is close to linear.} However, both the inductive bias of transformers and the dynamics of the microbiome are not understood well enough yet to draw any final conclusions from these intuitive arguments.  Therefore, an empirical compression with transformers would be highly interesting future work.

\textbf{SSM.} State space models (SSMs) \citep{SSMgu2021efficiently} are conceptually quite similar to NJODEs. Modern SSM techniques offer impressive benefits in terms of computation costs for large-scale problems. However, the computational costs are not the bottleneck for our setting. It might be interesting future work to compare NJODEs against SSMs, while we believe that NJODEs can deal better with small-scale training data with very irregular observations. SSMs would be particularly promising if we had access to the data of millions of patients with thousands of measurements per patient rather than hundreds of patients with tens of observations per patient.

\section{Discussion of Causality}\label{appendix:causality}
From a causal perspective, the reason (e.g., a disease) for the antibiotic treatment can be a relevant confounder influencing the gut microbiome both pre- and post-antibiotic exposure. The disease requiring antibiotic treatment could additionally be causing a decrease in the gut microbial diversity, making the comparison of anomaly scores before and after antibiotic exposure appear less significant than it actually is (Case 1). In this case, even if our method correctly identifies two consecutive anomalies (one from the disease, one from the antibiotic), we might not see a significant \emph{relative} score increase after the antibiotic exposure compared to the score before in \Cref{fig:score}b (Case 1a). Alternatively, the anomalous effect of an antibiotic on an already perturbed microbiome could be weaker than the effect on a healthy microbiome (Case 1b), making our anomaly quantification not generalizable to a healthy microbiome. From a causal theory perspective, correlations between antibiotic exposure and microbiome anomalies may exist without direct causal effects (Case 2). However, this case lacks empirical support in the current microbiome literature. 

Higher temporal resolution of antibiotic administration data, increased sampling frequency of the microbiome, and larger cohorts stratified by treatment reasons would allow the presented anomaly detection framework to further delineate the  antibiotic effects enabling causal inference of antibiotic exposure, underlying pathology, and microbiome perturbations. 

\newpage

\FloatBarrier

% Cohort information
\begin{table}[hp]
    \centering
    \caption{Cohort metadata counts}
    \label{supptab:cohort_counts}
    \begin{tabular}{l l r}
        \toprule
        \textbf{Characteristic} & \textbf{Category} & \textbf{Number of infants} \\
        \midrule
        Birth Mode              & Vaginal           & 254                        \\
                                & C-section         & 27                         \\
        \midrule
        Sex                     & Female            & 128                        \\
                                & Male              & 153                        \\
        \midrule
        Location                & Finland           & 132                        \\
                                & Estonia           & 77                         \\
                                & Russia            & 72                         \\
        \bottomrule
    \end{tabular}
\end{table}

% Data dictionary
\begin{table}[hp]
    \centering
    \renewcommand{\arraystretch}{1.5}
    \caption{Data dictionary}
    \label{supptab:datadict}
    \begin{tabular}{|l|}
        \hline
        \textbf{File}                                                                                                                                     \\
        \hline
        \href{https://github.com/adamovanja/anomaly_microbiome_data_processing/blob/main/data/final/data_dictionary_anomaly_v20240806.xlsx}{link to file} \\
        \hline
    \end{tabular}
\end{table}

% Grouping of antibiotics
\begin{table}[hp]
    \centering
    \renewcommand{\arraystretch}{1.5}
    \caption{Grouping of antibiotics}
    \label{supptab:abxinfo}
    \begin{tabular}{|l|}
        \hline
        \textbf{File}                                                                                                                   \\
        \hline
        \href{https://github.com/adamovanja/anomaly_microbiome_data_processing/blob/main/data/raw/clinical_maps_abx.xlsx}{link to file} \\
        \hline
    \end{tabular}
\end{table}

%TC:endignore
\end{document}